\documentclass[12pt,a4paper,final]{iopart}

\usepackage{iopams} 
\usepackage{wrapfig} 
\usepackage[breaklinks=true,colorlinks=true,linkcolor=blue,urlcolor=blue,citecolor=blue]{hyperref}

\expandafter\let\csname equation*\endcsname\relax
\expandafter\let\csname endequation*\endcsname\relax
\usepackage{amsmath}
\usepackage{amssymb,amsfonts}
\usepackage{mathtools}
\usepackage{latexsym}
\usepackage{cite}
\usepackage{setspace}
\usepackage{listings}
\usepackage{color} %red, green, blue, yellow, cyan, magenta, black, white
\definecolor{mygreen}{RGB}{28,172,0} % color values Red, Green, Blue
\definecolor{mylilas}{RGB}{170,55,241}
\definecolor{mybrown}{RGB}{165,42,42}
\definecolor{mypink}{RGB}{255,20,147}
\definecolor{darkGreen}{RGB}{34,139,34}

\begin{document}

% \lstset{language=Matlab,%
%     breaklines=true,%
%     morekeywords={matlab2tikz},
%     keywordstyle=\color{blue},%
%     identifierstyle=\color{black},%
%     stringstyle=\color{mylilas},
%     commentstyle=\color{mygreen},%
%     showstringspaces=false,%without this there will be a symbol in the places where there is a space
%     numbers=left,%
%     numberstyle={\tiny \color{darkGreen}},% size of the numbers
%     numbersep=9pt, % this defines how far the numbers are from the text
%     emph=[1]{function,for,if,else,end,break},emphstyle=[1]\color{cyan},
%     emph=[2]{cayleyMatrices,qStruct,quaternion,bloch,blochSp,movingFrames,parallelTransport,makeFigure1,make2sphere,askContinue,makeFiberPlot,makeDarboux,bundleSpace,closedPathFigure},emphstyle=[2]\color{mybrown},%some words to emphasise
%     %emph=[2]{word1,word2}, emphstyle=[2]{style},    
% }

%% -------------------Alternative commands-------------------------

\newcommand{\ket}[1]{\vert#1\rangle}
\newcommand{\tket}[1]{\vert{\boldsymbol#1}\rangle}
\newcommand{\tbra}[1]{\langle{\boldsymbol#1}\vert}
\newcommand{\bra}[1]{\langle#1\vert}
\newcommand{\braket}[2]{\langle#1\vert#2\rangle}
\newcommand{\ketbra}[2]{\vert#1\rangle\langle#2\vert}
\newcommand{\cub}[1]{\left(#1\right)}
\newcommand{\sqb}[1]{\left[#1\right]}
\newcommand{\pli}[1]{\hat{\sigma}_{#1}}
\newcommand{\rvec}[1]{\overset{\shortleftarrow}{#1}}
\newcommand{\uv}[1]{\vec{x}_{#1}} % for unit vector
\newcommand{\dv}[1]{\underset{\vec{}\hspace{0.025cm}}{x}^{#1}} % Dual Vector
\newcommand{\uve}[1]{\vec{e}_{#1}} % for unit vector
\newcommand{\duve}[1]{\dot{\vec{e}}_{#1}} % for unit vector
\newcommand{\dve}[1]{\underset{\vec{}\hspace{0.025cm}}{e}^{#1}} % Dual Vector
\newcommand{\ddve}[1]{\underset{\vec{}\hspace{0.025cm}}{\dot{e}}^{#1}} % Dual Vector
\newcommand{\aff}[3]{\Gamma^{#1}_{\;\;#2#3}}
\newcommand{\pd}[2]{\frac{\partial #1}{\partial #2}} % for partial derivatives
\newcommand{\quat}[1]{\hat{\sigma}_{#1}}
\newcommand{\cay}[1]{\hat{\rho}_{#1}}
\newcommand{\suat}[1]{\hat{\gamma}_{#1}}

%% -------------------Alternative commands-------------------------

\title{Quaternions, Spinors and the Hopf Fibration: \newline\textit{ Hidden Variables in Classical Mechanics}\qquad}

\author{Brian O'Sullivan}

%\address{Valeo Vision Systems, Dunmore Road Ind. Est., Tuam, Co. Galway. \newline 
%	School of Computer Science, National University of Ireland, Galway.}

\begin{abstract}
Rotations in 3 dimensional space are equally described by the SU(2) and SO(3) groups. These are isomorphic groups that generate the same 3D kinematics using different algebraic structures of the unit quaternion. The Hopf Fibration is a projection between the hypersphere $\mathbb{S}^3$ of the quaternion in 4D space, and the unit sphere $\mathbb{S}^2$ in 3D space. Great circles in $\mathbb{S}^3$ are mapped to points in $\mathbb{S}^2$ via the 6 Hopf maps, and are illustrated via the stereographic projection. The higher and lower dimensional spaces are connected via the $\mathbb{S}^1$ fibre bundle which consists of the global, geometric and dynamic phases. The global phase is quantized in integer multiples of $2\pi$ and presents itself as a natural hidden variable of Classical Mechanics. 
\end{abstract}

Hidden variable theories of Quantum Mechanics have long been a topical theme due to the demonstrable non-locality and indeterminacy of the science. 
Quantum Mechanics was born at the turn of the 20th century when technology became sufficiently advanced so as to allow experiments on the smallest scales. The properties of light and atoms being measured were unable to be accounted for by the Classical Mechanics of that era, and so it became apparent early on that a more sophisticated theory of the fundamental processes is needed. In the absence of such a theory, many properties of light, atoms and electrons were catalogued while an underlying mathematical basis to justify their presence was unknown. In this vacuum of rational explanations for observed phenomena one axiom within the Quantum theory has risen to the fore - attaining popular account for the unexplainable - this axiom is the Born rule. Also known as the Copenhagen Interpretation of Quantum Mechanics, it states that the square modulus of the complex numbers describes the probable result of an experimental measure. Furthermore, that the fundamental particles exist in a superposition of probable states prior to measurement, and only following a measurement does the state, or wave-function collapse to provide the measured result.

This article takes the view that the Born rule axiom is inherently incorrect, and is founded on an improper understanding of the underlying mathematics. In addition it is here demonstrated that hidden variables do exist, in the sense that phenomenological effects observed at the smallest scales are signatures measured from higher dimensional spaces. A proper and complete mathematical theory of these higher dimensional spaces is required. 
These statements are justified in this article in two parts.
\begin{itemize}
	\item[$(i)$] In the opening section of this article it is shown that the SU(2) and SO(3) groups are isomorphic. The kinematics of 3-dimensional space, which is typically described using the SO(3) picture, is equally represented in the SU(2) picture. This is understood as the quaternion is foundational structure of both pictures. Quaternion rotations in SU(2) are identical to quaternion rotations in SO(3). 
	
	In the literature of Quantum Mechanics the SU(2) group is a common feature, while references to the quaternion remain absent - as the common treatment of this geometric object is the application of the Born rule to the qubit. Consequently it is easily demonstrated there is a widespread misunderstanding of quaternion algebra in modern Quantum Mechanics.
	\item[$(ii)$] The quaternion is a 4 dimensional mathematical object which describes 3D rotations in full generality. As a consequence of the dimensional mismatch, information from the 4D space is both lost and retained. This is eloquently captured in the Hopf Fibration\cite{Lyons:03}\cite{Ovsienko:15}, as it is demonstrated that 
	\begin{enumerate}
		\item[1)] Information is lost: Great circles in $\mathbb{S}^3$ correspond to points in $\mathbb{S}^2$, and 
		\item[2)] Information is retained: The signature of the 4$^{th}$ dimension (aka Hidden Variables) is encoded in the 3D kinematics through the global, geometric and dynamic phases of the $\mathbb{S}^1$ fibre bundle.  
	\end{enumerate}
	The hidden variables of the $\mathbb{S}^1$ fibre bundle are implicitly defined through the 3-dimensional kinematics, and are demonstrated in due course. 
\end{itemize}
These are the principle points of this article. This article is an extension of the results presented in \cite{Wharton:15} which showed that the global phase of the qubit is a natural hidden variable, concealed in the Bloch sphere dynamics via the $\mathbb{S}^3\mapsto\mathbb{S}^2$ map of the Hopf Fibration. We account for all 6 maps of the Hopf Fibration, present the $\mathbb{S}^1$ fibre bundle in its closed form, and demonstrate the quantization of the global phase for closed paths on $\mathbb{S}^2$.

%The misalignment between the Quantum theory and all other Classical theories of physics, including Relativity theory, can be surmised in two parts. The first is $(i)$ the use of the Born rule in Quantum Mechanics. This axiom is not found in any other discipline of Physics, and it's use in the Quantum theory is addressed over the course of this article. The second aspect of the Quantum theory which conflicts with Classical theories of Physics is $(ii)$ the definition of the geometric phase. In Classical Physics, the geometric phase is defined as the angular change of a 2-dimensional tangent vector, following the completion of a closed loop path on a 3-dimensional surface. The tangent vector exists in the 2D tangent plane of a 3D surface, and it's motion is along the path is correctly described using the equation of parallel transport, section~\ref{sec:hopf}.

This article is constructed as a counter argument to the Born rule, using the hidden variables of Classical Mechanics as a working example. In the opening section it is shown that the special unitary group of $2\times2$ matrices SU(2), which is the fundamental group of Quantum Mechanics, is isomorphic to SO(3) - the special orthogonal group of $3\times3$ matrices. All rotations in SO(3) are equally described by SU(2), as the governing equations of Classical Mechanics can be easily recast in the SU(2) picture. In section~\ref{sec:hopf}, the 6 maps of the Hopf Fibration from the hypersphere $\mathbb{S}^3$ in 4D space, to the Bloch sphere $\mathbb{S}^2$ in 3D space, are defined, and illustrated via the stereographic projection. The global, geometric and dynamic phases of the $\mathbb{S}^1$ fibre bundle are presented in their closed form. A numerical example is provided to show the global phase is quantized for all closed paths of the Bloch sphere. The global phase is a measure of the total anholonomy of the rotation, and determines whether a full rotation in $\mathbb{S}^2$ correspond to a full or half rotation in $\mathbb{S}^3$. This is a property characteristic of the spin-half particles, and suggests it may be appropriate for understanding the nature of particle spin. In section~\ref{sec:spinor} an account of the spin-half particles offered by modern Quantum Mechanics is detailed. This is the Born rule applied to the qubit (the $\mathbb{C}^2$ spinor), and is representative of the widespread misunderstandings of quaternion algebra in Quantum Mechanics.

\section{Quaternions}
\label{sec:quaternion}

The quaternion was discovered as an algebra in 1843 by William Rowan Hamilton \cite{Hamilton:44},  
generalizing the description of 2-dimensional rotations by the complex numbers $\mathbb{C}$, to describe 3-dimensional rotations in a natural way \cite[{\it ch} 11]{Penrose:04}. The quaternions are a 4-dimensional {`complex'} number which describe rotations in 3-dimensions, in full generality \cite{Kuipers:00}. Containing 1 {`real'} and 3 {`complex'} components, the quaternions are isomorphic to vectors in $\mathbb{R}^4$ in the same way that the complex numbers are isomorphic to vectors in $\mathbb{R}^2$. While the complex numbers $\mathbb{C}$ describe rotations in 2-dimensions, the quaternions $\mathbb{C}^2$ describe rotations in both 4-dimensions and 3-dimensions \cite{Thomas:14}. 

The basis matrices of the quaternion in the SU(2) representation are referred to in this article as the \textit{Cayley matrices}, they are defined: 
\begin{equation}
	\label{eq:cayleyBasis}
	\quat{i}\;\equiv\; 
	\begin{pmatrix} 
		i & 0 \\ 
		0 & -i 
	\end{pmatrix}\qquad
	\quat{j}\;\equiv\; 
	\begin{pmatrix} 
		0 & 1 \\ 
		-1 & 0 
	\end{pmatrix}\qquad
	\quat{k}\;\equiv\; 
	\begin{pmatrix} 
		0 & i \\ 
		i & 0 
	\end{pmatrix}
\end{equation}
$\quat{1}$ is the $2\times2$ identity matrix, and $i=\sqrt{-1}$.
The Cayley matrices follow the multiplicative law of the quaternion algebra:
\begin{equation} 
	\label{eq:cayleyLaw}
	\quat{i}^2=\quat{j}^2=\quat{k}^2=\quat{i}\quat{j}\quat{k}=-\quat{1}
\end{equation}
The unit quaternion, with $$a^2+b^2+c^2+d^2=1$$ 
describes the hypersphere $\mathbb{S}^3$ which is a subspace of $\mathbb{R}^4$.
The quaternion  $\hat{U}_{\hat{\sigma}}\in\mathbb{S}^3\subset\mathbb{R}^4$ is expanded in the SU(2) Cayley basis as
\begin{equation}
\label{eq:quat}
\hat{U}_{\hat{\sigma}}\;=\;a\quat{1} + b\quat{i} + c\quat{j} + d\quat{k} \;=\; 
\begin{pmatrix}
a + ib & c + id \\ 
-c + id & a-ib
\end{pmatrix}		
\end{equation}
$\hat{U}_{\hat{\sigma}}^\dagger$ is the transpose conjugate of the SU(2) matrix.
\begin{equation}
\label{eq:quatD}
\hat{U}_{\hat{\sigma}}^\dagger\;=\;a\quat{1} - b\quat{i} - c\quat{j} - d\quat{k} \;=\; 
\begin{pmatrix}
a - ib & - c - id \\ 
c - id & a + ib
\end{pmatrix}		
\end{equation}
and $\hat{U}_{\hat{\sigma}}^\dagger\hat{U}_{\hat{\sigma}}=\hat{U}_{\hat{\sigma}}\hat{U}_{\hat{\sigma}}^\dagger=\quat{1}$.

The unit quaternions describe rotations in 3-dimensional space. 
The 3-vector $\vec{\mathcal{R}}$ is expanded in $\mathbb{R}^3$ using the SU(2) basis via
\begin{equation}
\label{eq:vecR}
\hat{\mathcal{R}}_{\hat{\sigma}}\;=\; 
\mathcal{R}^i\quat{i} + 
\mathcal{R}^j\quat{j} + 
\mathcal{R}^k\quat{k} \;=\;
\begin{pmatrix}
i\mathcal{R}^i & \mathcal{R}^j + i\mathcal{R}^k \\ 
-\mathcal{R}^j + i\mathcal{R}^k & -i\mathcal{R}^i
\end{pmatrix}		
\end{equation}
The vector $\hat{\mathcal{R}}_{\hat{\sigma}}$ is rotated to a new position $\hat{\mathcal{R}}_{\hat{\sigma}}'$ via the unit quaternion $\hat{U}_{\hat{\sigma}}$ as
\begin{equation}
\label{eq:rotateRsu2}
\hat{\mathcal{R}}_{\hat{\sigma}}'\;=\;\hat{U}_{\hat{\sigma}} \;\hat{\mathcal{R}}_{\hat{\sigma}} \; \hat{U}_{\hat{\sigma}}^\dagger 
\end{equation}
Similarly the rotation of a quaternion $\hat{P}_{\hat{\sigma}}\in\mathbb{S}^3\subset\mathbb{R}^4$ to a new position $\hat{P}_{\hat{\sigma}}'$ is described by 
$\hat{P}_{\hat{\sigma}}'=\hat{U}_{\hat{\sigma}}\hat{P}_{\hat{\sigma}}\hat{U}_{\hat{\sigma}}^\dagger $.
Equation \eqref{eq:rotateRsu2} describes the basics of rotations in $\mathbb{R}^3$ using unit quaternions in the SU(2) picture.
A greater understanding of the relationship between rotations in $\mathbb{R}^3$ using the SU(2) picture, and the SO(3) picture, is found by first moving to the $4\times4$ matrix representations of the quaternion.

\subsection*{Quaternions as $4\times4$ matrices}

Quaternions can be represented as $4\times4$ matrices \cite{Horn:86}. All that is required is the $4\times4$ basis matrices satisfy the law of quaternion multiplication quoted in equation~\eqref{eq:cayleyLaw}. There are 2 equivalent representations in the $4\times4$ picture, here referred to as the \textit{left Cayley} and \textit{right Cayley} matrices. The left Cayley matrices are defined
\begin{equation}
\label{eq:cayleyL}
\hat{l}_i\;=\; 
\begin{pmatrix}
0 & -1 & 0 & 0 \\
1 & 0 & 0 & 0 \\
0 & 0 & 0 & -1 \\
0 & 0 & 1 & 0
\end{pmatrix}\;\;\;\;
\hat{l}_j\;=\; 
\begin{pmatrix}
0 & 0 & -1 & 0 \\
0 & 0 & 0 & 1 \\
1 & 0 & 0 & 0 \\
0 & -1 & 0 & 0
\end{pmatrix}\;\;\;\;
\hat{l}_k\;=\; 
\begin{pmatrix}
0 & 0 & 0 & -1 \\
0 & 0 & -1 & 0 \\
0 & 1 & 0 & 0 \\
1 & 0 & 0 & 0
\end{pmatrix}
\end{equation}
which satisfy the relation 
$$\hat{l}_i^2\;=\;\hat{l}_j^2\;=\;\hat{l}_k^2\;=\;\hat{l}_i\hat{l}_j\hat{l}_k\;=\;-\hat{l}_1$$
The right Cayley matrices are defined
\begin{equation}
\label{eq:cayleyR}
\hat{r}_i\;=\; 
\begin{pmatrix}
0 & 1 & 0 & 0 \\
-1 & 0 & 0 & 0 \\
0 & 0 & 0 & -1 \\
0 & 0 & 1 & 0
\end{pmatrix}\;\;\;\;
\hat{r}_j\;=\; 
\begin{pmatrix}
0 & 0 & 1 & 0 \\
0 & 0 & 0 & 1 \\
-1 & 0 & 0 & 0 \\
0 & -1 & 0 & 0
\end{pmatrix}\;\;\;\;
\hat{r}_k\;=\; 
\begin{pmatrix}
0 & 0 & 0 & 1 \\
0 & 0 & -1 & 0 \\
0 & 1 & 0 & 0 \\
-1 & 0 & 0 & 0
\end{pmatrix}
\end{equation}
which satisfy the relation 
$$\hat{r}_i^2\;=\;\hat{r}_j^2\;=\;\hat{r}_k^2\;=\;\hat{r}_i\hat{r}_j\hat{r}_k\;=\;-\hat{r}_1$$
and $\hat{l}_1,\hat{r}_1$ are the $4\times4$ identity matrices.
The quaternion $\hat{U}$ is expanded in the left and right Cayley bases respectively as
\begin{align}\label{eq:qLeft}
\hat{U}_{\hat{l}}\;&=\;a\;\hat{l}_1+b\;\hat{l}_i+c\;\hat{l}_j+d\;\hat{l}_k\;=\;
\begin{pmatrix}
a & -b & -c & -d \\
b & a & -d & c \\
c & d & a & -b \\
d & -c & b & a
\end{pmatrix}\\ 
\label{eq:qRight}
\hat{U}_{\hat{r}}\;&=\;a\;\hat{r}_1+b\;\hat{r}_i+c\;\hat{r}_j+d\;\hat{r}_k\;=\;
\begin{pmatrix}
a & b & c & d \\
-b & a & -d & c \\
-c & d & a & -b \\
-d & -c & b & a
\end{pmatrix} 
\end{align}
The 3-vector $\hat{\mathcal{R}}$ is expanded in the left Cayley and right Cayley basis respectively as
\begin{align*}
\hat{\mathcal{R}}_{\hat{l}} \;&=\; 
\mathcal{R}^i\hat{l}_i+\mathcal{R}^j\hat{l}_j+\mathcal{R}^k\hat{l}_k\\
\hat{\mathcal{R}}_{\hat{r}} \;&=\; 
\mathcal{R}^i\hat{r}_i+\mathcal{R}^j\hat{r}_j+\mathcal{R}^k\hat{r}_k
\end{align*}
and is rotated to a new position $\hat{\mathcal{R}}'$ in the left and right Cayley basis respectively as
\begin{equation}
\label{eq:rotateRso4}
\hat{\mathcal{R}}_{\hat{l}}' \;=\; 
\hat{U}_{\hat{l}} \;\hat{\mathcal{R}}_{\hat{l}} \; \hat{U}^t_{\hat{l}}  \qquad\qquad\qquad
\hat{\mathcal{R}}_{\hat{r}}' \;=\; 
\hat{U}_{\hat{r}} \;\hat{\mathcal{R}}_{\hat{r}} \; \hat{U}^t_{\hat{r}}
\end{equation}
The result of the rotation in either basis is the same.\footnote[7]{
The quaternions expanded in the left and right Cayley bases commutate:
$[\hat{U}_{\hat{l}},\hat{U}_{\hat{r}}]=
[\hat{U}_{\hat{l}},\hat{U}_{\hat{r}}^t]=
[\hat{U}_{\hat{l}}^t,\hat{U}_{\hat{r}}]=
[\hat{U}_{\hat{l}}^t,\hat{U}_{\hat{r}}^t]=\hat{0}$.
} 
Similarly the quaternion $\hat{P}$ expanded in the left or right Cayley bases is rotated via 
$\hat{P}_{\hat{l}}' = \hat{U}_{\hat{l}} \hat{P}_{\hat{l}} \hat{U}^t_{\hat{l}}$, and
$\hat{P}_{\hat{r}}' = \hat{U}_{\hat{r}} \hat{P}_{\hat{r}} \hat{U}^t_{\hat{r}}$.

\subsection*{Rotations in SO(3)}

The special orthogonal group of $3\times3$ matrices SO(3) is derived from the product of the quaternion expanded in both the left and right Cayley bases.
\begin{equation*}
\hat{U}_{\hat{l}}\hat{U}_{\hat{r}} \;=\;
\begin{pmatrix}
1 & 0 & 0 & 0 \\
0 & a^2+b^2-c^2-d^2 & 2\cub{bc-ad} & 2\cub{bd+ac} \\
0 & 2\cub{bc+ad} & a^2-b^2+c^2-d^2 & 2\cub{cd-ab} \\
0 & 2\cub{bd-ac} & 2\cub{cd+ab} & a^2-b^2-c^2+d^2
\end{pmatrix}
\end{equation*}
From the above we define the SO(3) rotation matrix.
\begin{equation}
\label{eq:so3}
\hat{U}\;\equiv\;
\begin{pmatrix}
a^2+b^2-c^2-d^2 & 2\cub{bc-ad} & 2\cub{bd+ac} \\
2\cub{bc+ad} & a^2-b^2+c^2-d^2 & 2\cub{cd-ab} \\
2\cub{bd-ac} & 2\cub{cd+ab} & a^2-b^2-c^2+d^2
\end{pmatrix}
\end{equation}
The bloch vector $\mathcal{\hat{R}}$ is expanded in the Lie algebra basis 
\begin{equation*}
\hat{\mathcal{R}}\;=\;
\mathcal{R}^i\hat{\pi}_i + \mathcal{R}^j\hat{\pi}_j + \mathcal{R}^k\hat{\pi}_k
\end{equation*}
where $(\hat{\pi}_i, \hat{\pi}_j, \hat{\pi}_k)$ are the Lie algebra matrices defined by
\begin{equation}
	\label{eq:lie}
	\hat{\pi}_i\;\equiv\; 
	\begin{pmatrix}
		0 & 0 & 0 \\ 
		0 & 0 & -1 \\ 
		0 & 1 & 0 
	\end{pmatrix}\qquad 
	\hat{\pi}_j\;\equiv\; 
	\begin{pmatrix}
		0 & 0 & 1 \\ 
		0 & 0 & 0 \\ 
		-1 & 0 & 0 
	\end{pmatrix}\qquad 
	\hat{\pi}_k\;\equiv\; 
	\begin{pmatrix}
		0 & -1 & 0 \\ 
		1 & 0 & 0 \\ 
		0 & 0 & 0 
	\end{pmatrix}
\end{equation}	
The bloch vector evolves from it's initial state via 
\begin{equation}
	\label{eq:vonneumannSO3_0}
	\hat{\mathcal{R}}\;=\;\hat{U}\;\hat{\mathcal{R}}(0)\;\hat{U}^t
\end{equation}
The cartesian frame $(\vec{\sigma}_i,\vec{\sigma}_j,\vec{\sigma}_k)$ of $\mathbb{R}^3$ is defined:
\begin{equation*}
\label{eq:cart}
\vec{\sigma}_i\;=\; 
\begin{pmatrix}
	1 \\ 0 \\ 0
\end{pmatrix}	\qquad\qquad
\vec{\sigma}_j\;=\; 
\begin{pmatrix}
	0 \\ 1 \\ 0
\end{pmatrix}	\qquad\qquad
\vec{\sigma}_k\;=\; 
\begin{pmatrix}
	0 \\ 0 \\ 1
\end{pmatrix}	
\end{equation*}
and the Bloch vector is expanded in vector form 
\begin{equation}
\label{eq:r3vec}
\mathcal{\vec{R}}\;=\;
\mathcal{R}^i\vec{\sigma}_i + 
\mathcal{R}^j\vec{\sigma}_j + 
\mathcal{R}^k\vec{\sigma}_k
\;=\; 
\begin{pmatrix}
	\mathcal{R}^i \\ 
	\mathcal{R}^j \\ 
	\mathcal{R}^k
\end{pmatrix} 
\end{equation}
Consequently the rotation of the vector $\mathcal{\vec{R}}\in\mathbb{R}^3$ to a new position $\mathcal{\vec{R}}'$ is described
\begin{equation}
	\label{eq:rotateRso3}
	\vec{\mathcal{R}}'\;=\;\hat{U}\;\mathcal{\vec{R}}
\end{equation}	
These results demonstrate that the rotation of the vector $\hat{\mathcal{R}}_{\hat{\sigma}}$ in the SU(2) picture of equation~\eqref{eq:rotateRsu2}, is equivalent to the rotation of the vector $\hat{\mathcal{R}}_{\hat{l}, \hat{r}}$ in the $4\times4$ Cayley basis of equation~\eqref{eq:rotateRso4}, which are both equivalent to the rotation of the vector $\mathcal{\hat{R}}$ in the Lie algebra basis of equation~\eqref{eq:vonneumannSO3_0}, and the rotation of the vector $\mathcal{\vec{R}}$ in the SO(3) picture of equation~\eqref{eq:rotateRso3}.

\subsection*{Equations of motion}

From the time dependent SU(2) quaternion
\begin{equation*}
	\hat{U}_{\hat{\sigma}}(t) \;\equiv\;  a(t)\quat{1} + b(t)\quat{i} + c(t)\quat{j} + d(t)\quat{k} 
\end{equation*}
define the SU(2) Hamiltonian  
\begin{equation}
\label{eq:ham}
\hat{\mathcal{H}}_{\hat{\sigma}}(t)\;\equiv\;\dot{\hat{U}}_{\hat{\sigma}}\hat{U}_{\hat{\sigma}}^\dagger \;=\; 
\frac{\mathcal{H}^i}{2}\quat{i} + \frac{\mathcal{H}^j}{2}\quat{j} + \frac{\mathcal{H}^k}{2}\quat{k}	
\;=\;\frac{1}{2}
\begin{pmatrix}
	i\mathcal{H}^i & \mathcal{H}^j + i\mathcal{H}^k \\ 
	-\mathcal{H}^j + i\mathcal{H}^k & -i\mathcal{H}^i
\end{pmatrix}	
\end{equation}
with
\begin{subequations}
\label{eq:hamiltonian}
\begin{align}
\mathcal{H}^i\;&=\;2\cub{c\dot{d}-\dot{c}d+a\dot{b}-\dot{a}b}\\
\mathcal{H}^j\;&=\;2\cub{\dot{b}d-b\dot{d}+a\dot{c}-\dot{a}c}\\
\mathcal{H}^k\;&=\;2\cub{b\dot{c}-\dot{b}c+a\dot{d}-\dot{a}d}
\end{align}
\end{subequations}
From these relations we develop the expression for the rotation of the Bloch vector from an initial position $\mathcal{\hat{R}}_{\hat{\sigma}}(0)$ to the position $\mathcal{\hat{R}}_{\hat{\sigma}}$ at time $t$. Rewriting equation~\eqref{eq:rotateRsu2}
\begin{equation}
	\hat{\mathcal{R}}_{\hat{\sigma}}\;=\;\hat{U}_{\hat{\sigma}} \;\hat{\mathcal{R}}_{\hat{\sigma}}(0) \; \hat{U}_{\hat{\sigma}}^\dagger 
\end{equation}
and taking the derivative of both sides we find
\begin{align*}
\dot{\hat{\mathcal{R}}}_{\hat{\sigma}}\;&=\;\dot{\hat{U}}_{\hat{\sigma}} \;\hat{\mathcal{R}}_{\hat{\sigma}}(0) \; \hat{U}_{\hat{\sigma}}^\dagger 
+ \hat{U}_{\hat{\sigma}} \;\hat{\mathcal{R}}_{\hat{\sigma}}(0)\; \dot{\hat{U}}_{\hat{\sigma}}^\dagger  \\ 
\dot{\hat{\mathcal{R}}}_{\hat{\sigma}}\;&=\;\cub{\dot{\hat{U}}_{\hat{\sigma}}  \hat{U}_{\hat{\sigma}}^\dagger }\cub{\hat{U}_{\hat{\sigma}}\;\hat{\mathcal{R}}_{\hat{\sigma}}(0)\; \hat{U}_{\hat{\sigma}}^\dagger }
+ \cub{\hat{U}_{\hat{\sigma}} \;\hat{\mathcal{R}}_{\hat{\sigma}}(0)\; \hat{U}_{\hat{\sigma}}^\dagger }\cub{\hat{U}_{\hat{\sigma}}\dot{\hat{U}}_{\hat{\sigma}}^\dagger} \\ 
\dot{\hat{\mathcal{R}}}_{\hat{\sigma}}\;&=\;\hat{\mathcal{H}}_{\hat{\sigma}}\hat{\mathcal{R}}_{\hat{\sigma}}
- \hat{\mathcal{R}}_{\hat{\sigma}} \hat{\mathcal{H}}_{\hat{\sigma}} 
\end{align*}
to arrive at the Von Neumann equation
\begin{equation}
	\label{eq:vonneumann}
\dot{\hat{\mathcal{R}}}_{\hat{\sigma}}\;=\;[\mathcal{H}_{\hat{\sigma}},\hat{\mathcal{R}}_{\hat{\sigma}}]
\end{equation}
We find analogous expressions in the $4\times4$ representation.
The Hamiltonian is expanded in the left and right Cayley bases as
\begin{align*}
\hat{\mathcal{H}}_{\hat{l}}\;&=\; \frac{\mathcal{H}^i}{2}\hat{l}_i + \frac{\mathcal{H}^j}{2}\hat{l}_j + \frac{\mathcal{H}^k}{2}\hat{l}_k	\;=\;\frac{1}{2}
\begin{pmatrix}
	0 & -\mathcal{H}^i & -\mathcal{H}^j & -\mathcal{H}^k \\
	\mathcal{H}^i & 0 & -\mathcal{H}^k & \mathcal{H}^j \\
	\mathcal{H}^j & \mathcal{H}^k & 0 & -\mathcal{H}^i \\
	\mathcal{H}^k & -\mathcal{H}^j & \mathcal{H}^i & 0
\end{pmatrix}\\ 
\hat{\mathcal{H}}_{\hat{r}}\;&=\;\frac{\mathcal{H}^i}{2}\hat{r}_i + \frac{\mathcal{H}^j}{2}\hat{r}_j + \frac{\mathcal{H}^k}{2}\hat{r}_k\;=\;
\frac{1}{2}
\begin{pmatrix}
	0 & \mathcal{H}^i & \mathcal{H}^j & \mathcal{H}^k \\
	-\mathcal{H}^i & 0 & -\mathcal{H}^k & \mathcal{H}^j \\
	-\mathcal{H}^j & \mathcal{H}^k & 0 & -\mathcal{H}^i \\
	-\mathcal{H}^k & -\mathcal{H}^j & \mathcal{H}^i & 0
\end{pmatrix} 
\end{align*}
and from the first derivative of equations~\eqref{eq:rotateRso4} we arrive again at the Von Neumann equations in the left and right Cayley bases.
\begin{equation}
	\label{eq:vonneumann4}
\dot{\hat{\mathcal{R}}}_{\hat{l}}\;=\;[\hat{\mathcal{H}}_{\hat{l}},\hat{\mathcal{R}}_{\hat{l}}]  \qquad\qquad\qquad
\dot{\hat{\mathcal{R}}}_{\hat{r}}\;=\;[\hat{\mathcal{H}}_{\hat{r}},\hat{\mathcal{R}}_{\hat{r}}]
\end{equation}

\subsection*{Equations of motion SO(3)}

In the SO(3) picture the Hamiltonian operator is defined 
\begin{equation*}
	\hat{\mathcal{H}}\;\equiv\; \dot{\hat{U}}\hat{U}^t \;=\; 
	\mathcal{H}^i \hat{\pi}_i + \mathcal{H}^j \hat{\pi}_j + \mathcal{H}^k \hat{\pi}_k \;=\; 
	\begin{pmatrix*}
		0 & -\mathcal{H}^k & \mathcal{H}^j \\ 
		\mathcal{H}^k & 0 & -\mathcal{H}^i \\ 
		-\mathcal{H}^j & \mathcal{H}^i & 0
	\end{pmatrix*}
\end{equation*}
Developing from equation~\eqref{eq:vonneumannSO3_0}
\begin{align*}
\dot{\hat{\mathcal{R}}}\;&=\;\dot{\hat{U}}\;\hat{\mathcal{R}}(0)\;\hat{U}^t
+\hat{U}\;\hat{\mathcal{R}}(0)\;\dot{\hat{U}}^t\\
\dot{\hat{\mathcal{R}}}\;&=\;
\cub{\dot{\hat{U}}\hat{U}^t}\cub{\hat{U}\;\hat{\mathcal{R}}(0)\;\hat{U}^t} 
+\cub{\hat{U}\;\hat{\mathcal{R}}(0)\;\hat{U}^t}\cub{\hat{U}\dot{\hat{U}}^t}
\end{align*}
to arrive at the SO(3) Von Neumann equation of motion 
\begin{equation}
\dot{\hat{\mathcal{R}}}\;=\;[\hat{\mathcal{H}}, \hat{\mathcal{R}}]
\end{equation}
The bloch vector evolves in time from it's initial state $\mathcal{\vec{R}}(0)$ according to equation \eqref{eq:rotateRso3}.
\begin{equation*}
	\vec{\mathcal{R}}\;=\;\hat{U}\;\mathcal{\vec{R}}(0)
\end{equation*}	
Take the time derivative of both sides
\begin{align*}
\dot{\vec{\mathcal{R}}}\;&=\;\dot{\hat{U}}\;\mathcal{\vec{R}}(0)
\;=\;\cub{\dot{\hat{U}}\hat{U}^t}\cub{\hat{U}\;\mathcal{\vec{R}}(0)}
\end{align*}
to find
\begin{equation*}
\dot{\vec{\mathcal{R}}}\;=\;\hat{\mathcal{H}}\;\vec{\mathcal{R}}
\end{equation*}	
Since the SO(3) Hamiltonian is a skew symmetric matrix we may express the equation of motion as the familiar vector equation from classical mechanics.
\begin{equation}
\label{eq:motionSO3}
\dot{\vec{\mathcal{R}}}\;=\;\vec{\mathcal{H}}\times\vec{\mathcal{R}}
\end{equation}
These calculations demonstrate the SO(3) classical mechanics equation of motion \eqref{eq:motionSO3} is equivalent to the familiar Von Neumann equation from Quantum Mechanics \eqref{eq:vonneumann}, \eqref{eq:vonneumann4}. Given that the SU(2) and SO(3) pictures are equivalent, we favour the SO(3) picture for the remainder of the text.

\section{The Hopf Fibration}
\label{sec:hopf}

%In the following we consider the quaternion $\hat{\Psi}$ which evolves from it's initial state $\hat{\Psi}_0$ according to the time dependent quaternion $\hat{U}$.
%\begin{equation}
%	\label{eq:Psi}
%\hat{\Psi}\;=\;\hat{U}\;\hat{\Psi}_0
%\qquad\qquad\qquad \qquad 
%\dot{\hat{\Psi}}\;=\;\hat{\mathcal{H}}\;\hat{\Psi}
%\end{equation}
The quaternion $\hat{\Psi}$ describes the hypersphere $\mathbb{S}^3$ embedded in $\mathbb{R}^4$.
\begin{equation*}
\hat{\Psi}\;\in\;\mathbb{S}^3\;\subset\;\mathbb{C}^2,\;\mathbb{R}^4
\end{equation*}
The Hopf fibration is a projection between the 3-sphere $\mathbb{S}^3\subset\mathbb{R}^4$ and the 2-sphere $\mathbb{S}^2\subset\mathbb{R}^3$. 
\begin{equation}
	\label{eq:hopf_fibration}
	\mathbb{S}^3 \; \xmapsto{\mathbb{S}^1} \; \mathbb{S}^2
\end{equation}
The $\mathbb{S}^3$ and $\mathbb{S}^2$ spaces are respectively the total space and base space, and are connected by the $\mathbb{S}^1$ fibre bundle which is the unit circle, detailed in section~\ref{sec:s1bundle}.
For a given quaternion $\hat{\Psi}$ there are 6 possible mappings from $\mathbb{S}^3\mapsto\mathbb{S}^2$, given by
\begin{subequations}
\label{eq:hopf_project}
\begin{align}
\hat{\mathcal{I}}_l\;=\;\hat{\Psi}^t\hat{\pi}_i\hat{\Psi}
\qquad\qquad\;\;
\hat{\mathcal{J}}_l\;&=\;\hat{\Psi}^t\hat{\pi}_j\hat{\Psi}
\qquad\qquad\;
\hat{\mathcal{K}}_l\;=\;\hat{\Psi}^t\hat{\pi}_k\hat{\Psi} \\ 
\hat{\mathcal{I}}_r\;=\;\hat{\Psi}\hat{\pi}_i\hat{\Psi}^t
\;\qquad\qquad
\hat{\mathcal{J}}_r\;&=\;\hat{\Psi}\hat{\pi}_j\hat{\Psi}^t
\qquad\qquad\;
\hat{\mathcal{K}}_r\;=\;\hat{\Psi}\hat{\pi}_k\hat{\Psi}^t 
\end{align}
\end{subequations}
where the bloch vectors 
\begin{equation*}
	\hat{\mathcal{I}}_l,\;\hat{\mathcal{I}}_r,\;
	\hat{\mathcal{J}}_l,\;\hat{\mathcal{J}}_r,\;
	\hat{\mathcal{K}}_l,\;\hat{\mathcal{K}}_r
	\;\in\;\mathbb{S}^2\;\subset\;\mathbb{R}^3
\end{equation*}
The subscripts $`l'$ and $`r'$ refer to left and right respectively. 
Expand the unit quaternion $\hat{\Psi}$ in the SO(3) picture.
\begin{equation}
\label{eq:quat_so3}
\hat{\Psi}\;=\; 
\begin{pmatrix}
q_1^2+q_i^2-q_j^2-q_k^2 & 2\cub{q_iq_j-q_1q_k} & 2\cub{q_iq_k+q_1q_j} \\
2\cub{q_iq_j+q_1q_k} & q_1^2-q_i^2+q_j^2-q_k^2 & 2\cub{q_jq_k-q_1q_i} \\
2\cub{q_iq_k-q_1q_j} & 2\cub{q_jq_k+q_1q_i} & q_1^2-q_i^2-q_j^2+q_k^2
\end{pmatrix}
\end{equation}
Consequently we attribute the elements of the left and right Hopf projections \eqref{eq:hopf_project}
\begin{equation*}
\hat{\Psi}\;=\;
\begin{pmatrix}
	\mathcal{I}^i_{l} & \mathcal{I}^j_{l} & \mathcal{I}^k_{l} \\
	\mathcal{J}^i_{l} & \mathcal{J}^j_{l} & \mathcal{J}^k_{l} \\
	\mathcal{K}^i_{l} & \mathcal{K}^j_{l} & \mathcal{K}^k_{l} 
\end{pmatrix}
\qquad\qquad\qquad\qquad
\hat{\Psi}\;=\;
\begin{pmatrix}
\mathcal{I}^i_{r} & \mathcal{J}^i_{r} & \mathcal{K}^i_{r} \\
\mathcal{I}^j_{r} & \mathcal{J}^j_{r} & \mathcal{K}^j_{r} \\
\mathcal{I}^k_{r} & \mathcal{J}^k_{r} & \mathcal{K}^k_{r} 
\end{pmatrix}
\end{equation*}
Assign the Bloch vector $\hat{\mathcal{R}}$ to the projection of choice 
\begin{equation*}
	\hat{\mathcal{R}}\;\mapsto\;
	\hat{\mathcal{I}}_r,\;\hat{\mathcal{I}}_l,\;
	\hat{\mathcal{J}}_r,\;\hat{\mathcal{J}}_l,\;
	\hat{\mathcal{K}}_r,\;\hat{\mathcal{K}}_l
\end{equation*}
and the resulting vector is expanded in the Lie algebra basis
\begin{equation*}
	\hat{\mathcal{R}}\;=\;\mathcal{R}^i\hat{\pi}_i+\mathcal{R}^j\hat{\pi}_j+\mathcal{R}^k\hat{\pi}_k
\end{equation*}

\subsection{Stereographic projection of the Hopf map} 

In projections between dimensional spaces some information is lost, while other information is retained. 
In the Hopf projection from $\mathbb{S}^3\mapsto\mathbb{S}^2$, 
what is retained is the $\mathbb{S}^1$ fibre bundle, detailed in section~\ref{sec:s1bundle}, 
and what is lost in is ability to specify exactly what quaternion in $\mathbb{S}^3$ generates a point in $\mathbb{S}^2$.
A plurality of quaternions in $\mathbb{S}^3$ correspond to a single point in $\mathbb{S}^2$. 
This is well illustrated using the stereographic projection of the quaternion.
In vector form the quaternion is represented
\begin{equation*}
\vec{\Psi}\;=\;
\begin{pmatrix}
q_1 \\
q_i \\
q_j \\
q_k 
\end{pmatrix}	
\end{equation*}
and the stereographic projection of the quaternion, is the map
\begin{equation*}
	\mathbb{S}^3/(1,0,0,0)\;\mapsto\;\mathbb{R}^3
\end{equation*}
given by 
\begin{equation}
	\label{eq:stereo_project}
	(q_1,q_i,q_j,q_k)\;\mapsto\;\cub{\frac{q_i}{1-q_1},\frac{q_j}{1-q_1},\frac{q_k}{1-q_1}}
\end{equation}
This mapping is valid for all points except the singularity point $\vec{\Psi}=(1,0,0,0)^t$.
We now account for the 6 Hopf maps listed in equation~\eqref{eq:hopf_project}. 
To do so we account for the first projection in detail, and the remaining 5 projections all follow the same logic.

Rotate the quaternion $\vec{\Psi}$ in the left Cayley basis, through an angle $\varphi$ in the $\hat{l}_i$ axis. 
\begin{align*}
\vec{\Psi}'\;&=\;\exp\sqb{\frac{\varphi}{2}\hat{l}_i}\vec{\Psi}\\
\vec{\Psi}'\;&=\;
\begin{pmatrix*}
\cos\cub{\tfrac{\varphi}{2}} & -\sin\cub{\tfrac{\varphi}{2}} & 0 & 0 \\ 
\sin\cub{\tfrac{\varphi}{2}} & \cos\cub{\tfrac{\varphi}{2}} & 0 & 0 \\ 
0 & 0 & \cos\cub{\tfrac{\varphi}{2}} & -\sin\cub{\tfrac{\varphi}{2}} \\
0 & 0 & \sin\cub{\tfrac{\varphi}{2}} & \cos\cub{\tfrac{\varphi}{2}} 
\end{pmatrix*}
\begin{pmatrix}
q_1 \\
q_i \\
q_j \\
q_k 
\end{pmatrix}
\;=\; 
\begin{pmatrix}
	\cos\cub{\tfrac{\varphi}{2}}q_1-q_i\sin\cub{\tfrac{\varphi}{2}} \\ 
	\cos\cub{\tfrac{\varphi}{2}}q_i+q_1\sin\cub{\tfrac{\varphi}{2}} \\ 
	\cos\cub{\tfrac{\varphi}{2}}q_j-q_k\sin\cub{\tfrac{\varphi}{2}} \\ 
	\cos\cub{\tfrac{\varphi}{2}}q_k+q_j\sin\cub{\tfrac{\varphi}{2}}
\end{pmatrix}
\end{align*}
Now expressing $\vec{\Psi}'$ in the SO(3) picture:
\begin{equation*}
\hat{\Psi}'\;=\; 
\begin{pmatrix}
\mathcal{I}^i_{l} & \mathcal{I}^j_{l} & \mathcal{I}^k_{l} \\
\mathcal{J}^i_{l}\cos\cub{\varphi} - \mathcal{K}^i_{l}\sin\cub{\varphi} & 
\mathcal{J}^j_{l}\cos\cub{\varphi} - \mathcal{K}^j_{l}\sin\cub{\varphi} & 
\mathcal{J}^k_{l}\cos\cub{\varphi} - \mathcal{K}^k_{l}\sin\cub{\varphi} \\
\mathcal{K}^i_{l}\cos\cub{\varphi} + \mathcal{J}^i_{l}\sin\cub{\varphi} & 
\mathcal{K}^j_{l}\cos\cub{\varphi} + \mathcal{J}^j_{l}\sin\cub{\varphi} & 
\mathcal{K}^k_{l}\cos\cub{\varphi} + \mathcal{J}^k_{l}\sin\cub{\varphi}
\end{pmatrix}\\
\end{equation*}
where $\varphi\in[0,4\pi]$, and apply the appropriate Hopf map
\begin{equation*}
\hat{\mathcal{I}}_l\;=\;\hat{\Psi}^t\hat{\pi}_i\hat{\Psi}
\;=\;\hat{\Psi}'^t\hat{\pi}_i\hat{\Psi}'
\end{equation*}
Thus the set of quaternions described by $\hat{\Psi}'$ correspond to a single point in $\mathbb{S}^2$.
Similarly for the $j$ and $k$ projections:
\begin{align*}
\vec{\Psi}'\;=\;\exp\sqb{\frac{\varphi}{2}\hat{l}_j}\vec{\Psi}
\qquad\to\qquad&
\hat{\mathcal{J}}_l\;=\;\hat{\Psi}^t\hat{\pi}_j\hat{\Psi}
\;=\;\hat{\Psi}'^t\hat{\pi}_j\hat{\Psi}'\\
\vec{\Psi}'\;=\;\exp\sqb{\frac{\varphi}{2}\hat{l}_k}\vec{\Psi}
\qquad\to\qquad&
\hat{\mathcal{K}}_l\;=\;\hat{\Psi}^t\hat{\pi}_k\hat{\Psi}
\;=\;\hat{\Psi}'^t\hat{\pi}_k\hat{\Psi}'
\end{align*}	
and for the remaining 3 quaternion rotations in the right Cayley basis:
\begin{align*}
\vec{\Psi}'\;=\;\exp\sqb{\frac{\varphi}{2}\hat{r}_i}\vec{\Psi}
\qquad\to\qquad&
\hat{\mathcal{I}}_r\;=\;\hat{\Psi}\hat{\pi}_i\hat{\Psi}^t
\;=\;\hat{\Psi}'\hat{\pi}_i\hat{\Psi}'^t\\
\vec{\Psi}'\;=\;\exp\sqb{\frac{\varphi}{2}\hat{r}_j}\vec{\Psi}
\qquad\to\qquad&
\hat{\mathcal{J}}_r\;=\;\hat{\Psi}\hat{\pi}_j\hat{\Psi}^t
\;=\;\hat{\Psi}'\hat{\pi}_j\hat{\Psi}'^t\\
\vec{\Psi}'\;=\;\exp\sqb{\frac{\varphi}{2}\hat{r}_k}\vec{\Psi}
\qquad\to\qquad&
\hat{\mathcal{K}}_r\;=\;\hat{\Psi}\hat{\pi}_k\hat{\Psi}^t
\;=\;\hat{\Psi}'\hat{\pi}_k\hat{\Psi}'^t
\end{align*}	
\begin{figure}[t]
\begin{center} 
\includegraphics[width=\textwidth]{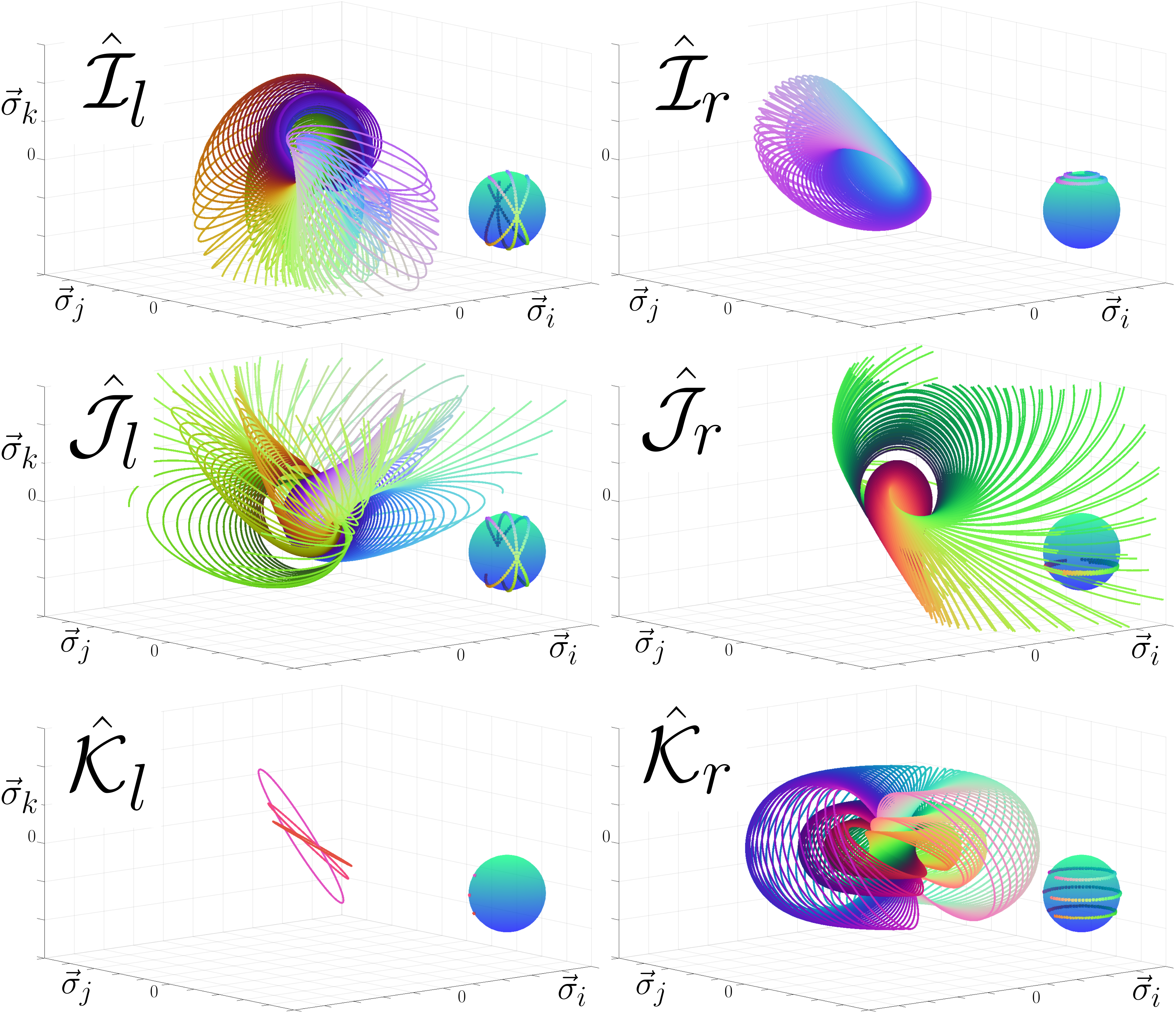}
\caption{Shown are the stereographic projections \eqref{eq:stereo_project} of the quaternion $\vec{\Psi}'$
		 according to the 6 Hopf maps of equation~\eqref{eq:hopf_project}. Matlab and python scripts used to generate the figures in this paper are found in the repository: \href{https://github.com/mo-geometry/hopf_fibration}{github.com/mo-geometry}}
\end{center}
\label{fig:fiberbundle} 
\end{figure}
To illustrate the stereographic projection of the Hopf map, we parametrize the quaternion via:
\begin{subequations}
	\label{eq:parametrize_q}
\begin{align*}
\textrm{SO(3) basis:}\qquad\qquad\qquad\;
\hat{\Psi}\;&=\; 
\exp\sqb{\phi\;\hat{\pi}_k}
\exp\sqb{\theta\;\hat{\pi}_j}
\exp\sqb{\eta\;\hat{\pi}_k}\\
\textrm{left Cayley basis:}\qquad\qquad\qquad
\hat{\Psi}_{\hat{l}}\;&=\; 
\exp\sqb{\frac{\phi}{2}\;\hat{l}_k}
\exp\sqb{\frac{\theta}{2}\;\hat{l}_j}
\exp\sqb{\frac{\eta}{2}\;\hat{l}_k}
\end{align*}
\end{subequations}
The left Cayley basis is quoted here as it is easier from a numerical point of view to extract the quaternion coefficients $(q_1,q_i,q_j,q_k)$. The great circles of $\mathbb{S}^3$ are illustrated in $\mathbb{R}^3$ via the stereographic projection \eqref{eq:stereo_project}, for each of the 6 Hopf maps of equation \eqref{eq:hopf_project}, as seen in figure~\ref{fig:fiberbundle}. The quaternions corresponding to each point are generated by creating an array of points with $\theta=[\tfrac{\pi}{3}, \tfrac{\pi}{2}, \tfrac{2\pi}{3}]$, and for each value of $\theta$, let $\phi\in[0, 1.7\pi]$ be divided into $60$ points with equal spacing, setting $\eta=\tfrac{\pi}{2}$. Element-wise the SO(3) quaternion is parametrized as
\begin{align*}
\hat{\Psi}\;&=\; 
\begin{pmatrix}
\textrm{c}\cub{\theta}\textrm{c}\cub{\phi}\textrm{c}\cub{\eta} - \textrm{s}\cub{\phi}\textrm{s}\cub{\eta} &
-\textrm{c}\cub{\theta}\textrm{c}\cub{\phi}\textrm{s}\cub{\eta} - \textrm{s}\cub{\phi}\textrm{c}\cub{\eta} &
\textrm{s}\cub{\theta}\textrm{c}\cub{\phi} 
\\
\textrm{c}\cub{\theta}\textrm{s}\cub{\phi}\textrm{c}\cub{\eta} + \textrm{c}\cub{\phi}\textrm{s}\cub{\eta} & 
-\textrm{c}\cub{\theta}\textrm{s}\cub{\phi}\textrm{s}\cub{\eta} + \textrm{c}\cub{\phi}\textrm{c}\cub{\eta} &  \textrm{s}\cub{\theta}\textrm{s}\cub{\phi}
\\
- \textrm{s}\cub{\theta}\textrm{c}\cub{\eta} & 
\textrm{s}\cub{\theta}\textrm{s}\cub{\eta} & 
\textrm{c}\cub{\theta} 
\end{pmatrix}
\end{align*}
with s$\cub{\bullet}$, c$\cub{\bullet}\;=\;\sin\cub{\bullet},\;\cos\cub{\bullet}$.

\subsection{The $\mathbb{S}^1$ fibre bundle}
\label{sec:s1bundle}

$\mathbb{S}^1$ fibre bundle connects the total space $\mathbb{S}^3$ and the subspace $\mathbb{S}^2$, and consists of the global, geometric and dynamic phases. The global phase is the sum of the geometric and dynamic phase. The dynamic phase is a measure of the total work over the path, and the geometric phase is the change in orientation of a parallel transported tangent vector. 
The geometric phase was originally studied in modern Quantum Mechanics in the context of adiabatically evolving quantum systems where it was acknowledged the global phase is the sum of the geometric and dynamic phases \cite{Berry:84}. Shortly thereafter, it was recognized that the global phase $\omega$ is a measure of the anholonomy of the $\mathbb{C}^2$ spinor's $\mathbb{S}^2$ path, and that the geometric phase $\gamma$ and dynamic phase $\xi$ constitute the elements of a fibre bundle \cite{Barry:83}. 
\begin{equation}
	\label{eq:fiber_bundle}
	\omega\;=\;\gamma+\xi 
\end{equation}
This fibre bundle is the unit circle $\mathbb{S}^1$ connecting the total space $\mathbb{S}^3$ and base space $\mathbb{S}^2$.
$$\mathbb{S}^3 \; \xmapsto{\mathbb{S}^1} \; \mathbb{S}^2$$
We proceed to define the parameters of the $\mathbb{S}^1$ fibre bundle using a Classical physics approach. 
The dynamic phase is defined as the total work 
\begin{equation}
\label{eq:dynamic_phase}
\xi\;=\;\int_0^tdt'\;\dot{\xi}(t')
\qquad\qquad\qquad\qquad
\dot{\xi}\;\equiv\;\mathcal{\vec{H}}\cdot \mathcal{\vec{R}}
\end{equation}
The geometric phase is defined as the angular change of a parallel transported 2D tangent vector, following the completion of a closed loop path on a 3D surface. Given that the angular change of the tangent vector is the quantity of interest, and not the precession relative to the tangent frame, the geometric phase is confined within in the range $\gamma\in[-\pi,\pi]$. 

The tangent vector exists in the tangent plane of the surface, which is mapped by a moving frame whose definition depends on the chosen coordinate system. We are free to choose any coordinate system in which to do this calculation,\footnote[7]{See \ref{app:geometric_phase} for some interesting graphical results of the differences between the full precession of the geometric phase in different coordinate systems.} and a natural choice is the Darboux tangent frame \cite{Jamiolkowski:04}. The Darboux moving frame consists of the unit velocity vector $\uve{v}$ which points along the direction of motion, and the surface bi-normal which is the normalized cross product of the surface normal and $\uve{v}$. The unit velocity vector is defined:
\begin{equation*}
	\uve{v}\;\equiv\;\frac{\mathcal{\vec{H}}\times\mathcal{\vec{R}}}{|\mathcal{\vec{H}}\times\mathcal{\vec{R}}|}
	\;=\;
	\frac{\mathcal{\vec{H}}\times\mathcal{\vec{R}}}
	{\sqrt{\mathcal{\vec{H}}\cdot\mathcal{\vec{H}} - \dot{\xi}^2}}
\end{equation*}
where we have made use of equation~\eqref{eq:motionSO3}. The surface normal of the unit sphere is the Bloch vector $\vec{\mathcal{R}}$. Therefore the bi-normal vector is defined
\begin{equation*}
	\uve{b}\;\equiv\;\frac{\mathcal{\vec{R}}\times\cub{\mathcal{\vec{H}}\times\mathcal{\vec{R}}}}
	{\left|\mathcal{\vec{R}}\times\cub{\mathcal{\vec{H}}\times\mathcal{\vec{R}}}\right|}
	\;=\;
	\frac{\mathcal{\vec{H}} - \mathcal{\vec{R}} \dot{\xi}}
	{\sqrt{\mathcal{\vec{H}}\cdot\mathcal{\vec{H}} - \dot{\xi}^2}}
\end{equation*}
The vectors $(\uve{v},\uve{b})$ form a moving frame whose origin is defined by $\vec{\mathcal{R}}$. A tangent vector $\mathcal{\vec{V}}$ is expanded in the moving frame via
\begin{equation*}
\mathcal{\vec{V}}\;=\;\mathcal{V}^v\uve{v}+\mathcal{V}^b\uve{b}
\end{equation*}
The vector is parallel transported along the path ascribed by $\vec{\mathcal{R}}$ according to the equation of parallel transport \cite{Hobson:06}.
\begin{equation*}
	\frac{D\vec{\mathcal{V}}}{Dt}\;=\;\dot{\vec{\mathcal{V}}}\cdot\uve{\lambda}\;=\;0
	\qquad\qquad \forall\;\; \uve{\lambda}\in[\uve{v},\uve{b}]
\end{equation*}
Developing we obtain the pair of simultaneous differential equations
\begin{equation*}
\frac{D\vec{\mathcal{V}}}{Dt}\;=\;
\begin{cases}
\dot{\mathcal{V}}^v + \dot{e}_{b}\cdot\uve{v} \;\mathcal{V}^b \;=\;0 \\ 
\dot{\mathcal{V}}^b + \dot{e}_{v}\cdot\uve{b} \;\mathcal{V}^v \;=\;0
\end{cases}     
\end{equation*}
The derivative of the geometric phase is defined $\dot{\gamma}\equiv\dot{e}_{v}\cdot\uve{b}=-\dot{e}_{b}\cdot\uve{v}$,
\begin{equation}\label{eq:geo_phase}
\gamma\;=\;\int_0^tdt'\;\dot{\gamma}(t')
\qquad\qquad\qquad\qquad
\dot{\gamma}\;\equiv\;
\frac{
\dot{\vec{\mathcal{H}}}\cdot\dot{\vec{\mathcal{R}}}
- \dot{\vec{\mathcal{R}}}\cdot\dot{\vec{\mathcal{R}}}\dot{\xi} }
{\mathcal{\vec{H}}\cdot\mathcal{\vec{H}} - \dot{\xi}^2}
\end{equation} 
\begin{figure}[t]
	\includegraphics[width=\textwidth]{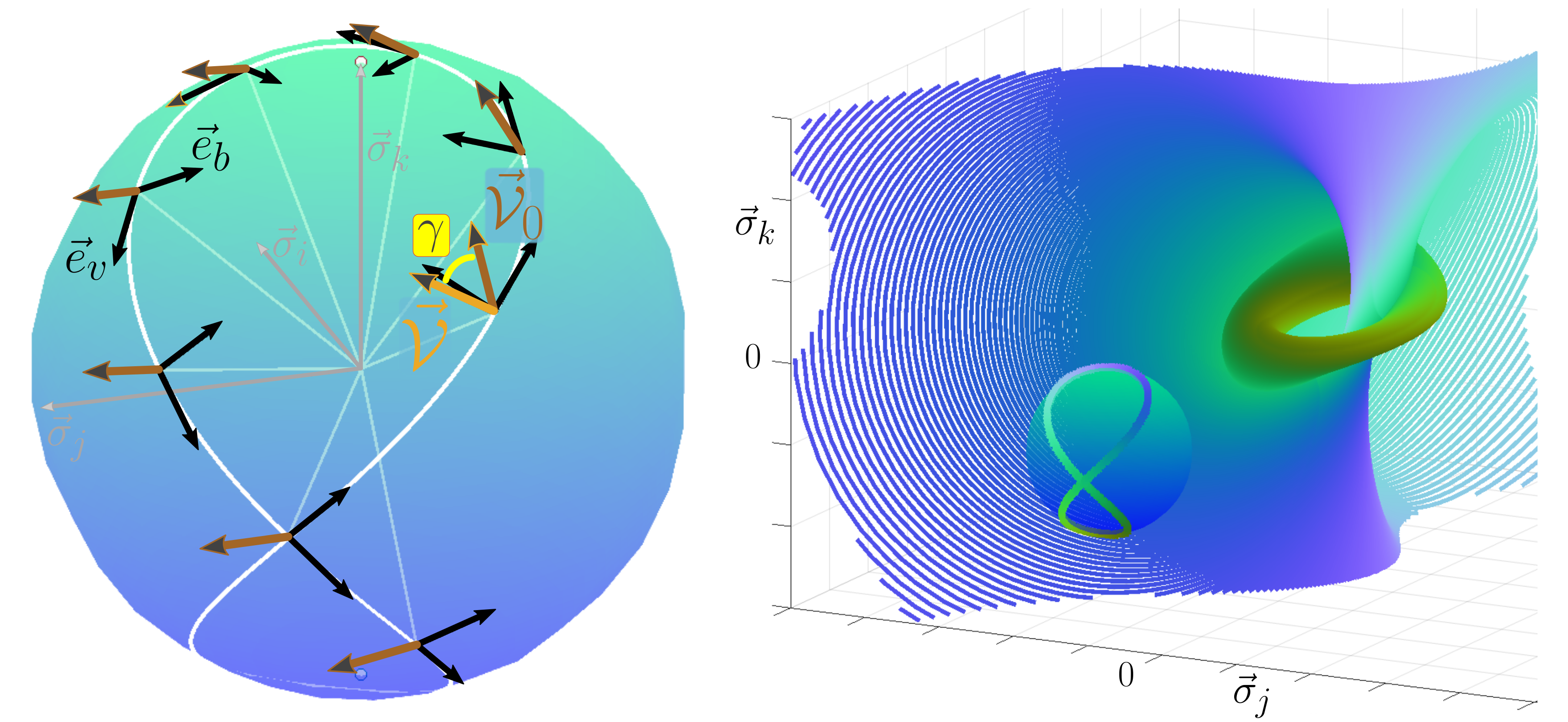}
	\caption{The Darboux tangent frame and a parallel transported tangent vector is illustrated on the Bloch sphere. The geometric phase (yellow) is highlighted as the change in orientation of the tangent vector when it returns to it's initial position. On the right is the Stereographic projection of the closed path under the $\hat{\mathcal{K}}_r$ Hopf mapping. } 
	\label{fig:bloch} 
\end{figure}
Express the equation of parallel transport in matrix form
\begin{equation*}
\begin{pmatrix}
\dot{\mathcal{V}}^v  \\ \dot{\mathcal{V}}^b 
\end{pmatrix}\;=\; 
\begin{pmatrix}
0 & \dot{\gamma} \\
-\dot{\gamma} & 0
\end{pmatrix}
\begin{pmatrix}
\mathcal{V}^v \\ \mathcal{V}^b 
\end{pmatrix}
\end{equation*}
the tangent vector evolves from it's initial state according to
\begin{equation*} 
\begin{pmatrix}
\mathcal{V}^v(t) \\ 
\mathcal{V}^b(t)
\end{pmatrix}
\;=\;
\begin{pmatrix}
 \cos\cub{\gamma} & \sin\cub{\gamma} \\
-\sin\cub{\gamma} & \cos\cub{\gamma} 
\end{pmatrix}
\begin{pmatrix}
\mathcal{V}^v(0) \\ 
\mathcal{V}^b(0)
\end{pmatrix}
\end{equation*}
The global phase is the sum of the geometric and dynamic phases \eqref{eq:fiber_bundle}, 
with the geometric phase confined to the range $\gamma\in[-\pi,\pi]$, and the dynamic phase is unbounded.
The geometric phase is invariant under the U(1) gauge as described in~\ref{app:U1gauge}.

To illustrate the geometric phase of a closed path in $\mathbb{S}^2$, and the associated stereographic projection of the path, we utilize the Bloch vector  
\begin{equation*}
	\hat{\mathcal{R}}\;=\;\hat{\Psi}\hat{\pi}_k\hat{\Psi}^t 
\end{equation*}
under the right Cayley Hopf map $\hat{\mathcal{K}}_r$, for the unitary,
\begin{align*}
	\textrm{SU(2) picture:}\qquad\qquad
	&\hat{U}_{\hat{\sigma}}(t)\;=\;\exp\sqb{\quat{k}\frac{t}{2}}\exp\sqb{\quat{j}\frac{t}{2}}\\
	\textrm{SO(3) picture:}\qquad\qquad
	&\;\hat{U}(t)\;=\;\exp\sqb{\hat{\pi}_{k} t}\exp\sqb{\hat{\pi}_{j}t}
\end{align*}
The quaternion $\hat{\Psi}=\hat{U}(t)\hat{\Psi}_0$, has an initial state $\hat{\Psi}_0$ parametrized according to \eqref{eq:parametrize_q}, with $(\theta_0,\phi_0,\eta_0)=(\tfrac{\pi}{3},1.056\pi,0)$. This generates the closed path shown in figure~\ref{fig:bloch}. The parallel transport of a tangent vector in the Darboux tangent frame is illustrated (left), alongside the great circles corresponding to the plurality of quaternions in $\mathbb{S}^3$ describing the $\mathbb{S}^2$ path under the $\hat{\mathcal{K}}_r$ stereographic projection (right).

\subsection{Numerical analysis of the $\mathbb{S}^1$ fibre bundle} 

\begin{figure}[t]
	\begin{center}
		\includegraphics[width=\textwidth]{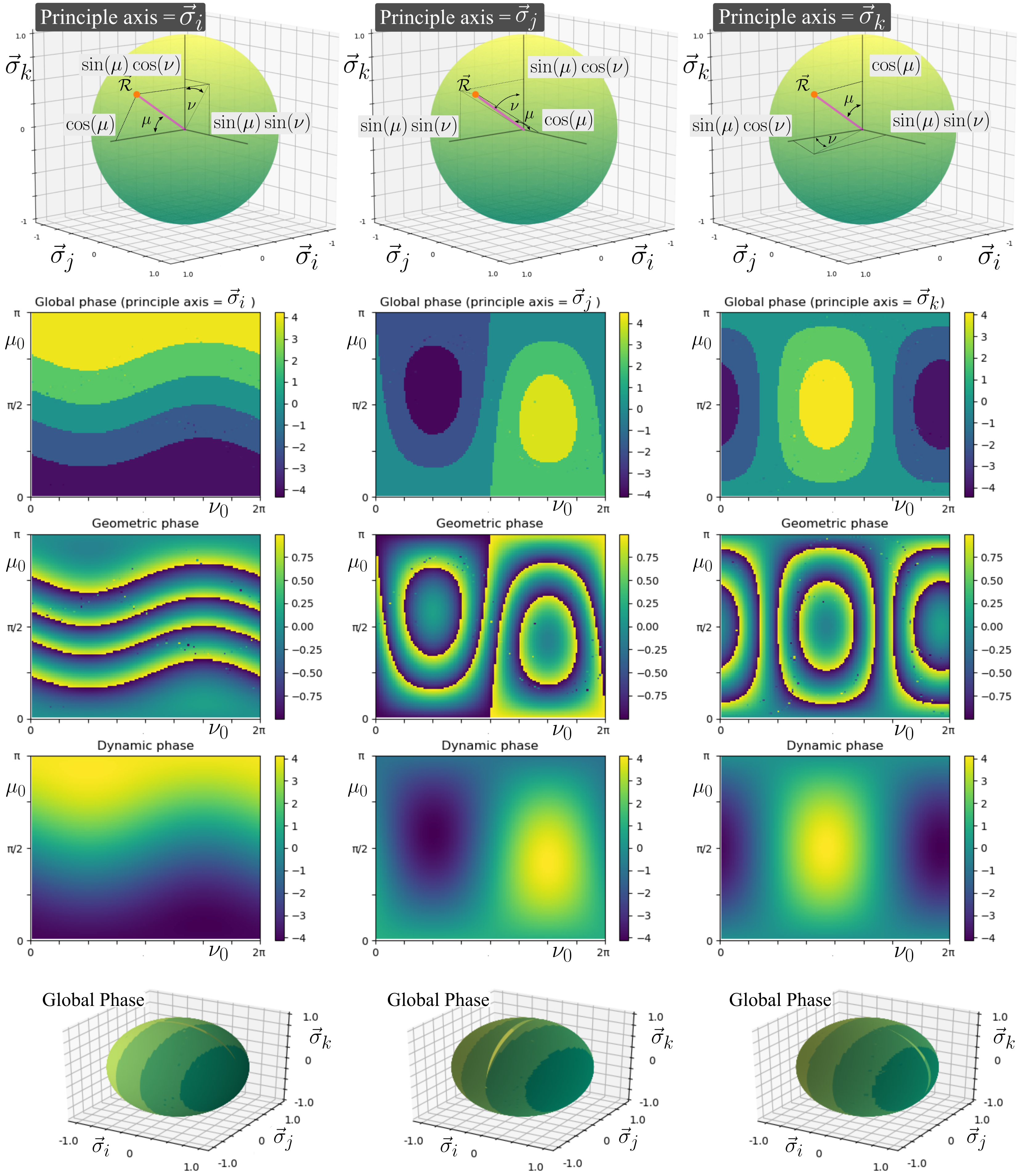}
		\caption{\textit{Top row:} Spherical polar coordinates, with principle axis $\vec{\sigma}_i, \vec{\sigma}_j, \vec{\sigma}_k$, respectively. \textit{Second row:} The global phase of the closed path for the unitary of equation~\eqref{eq:unitary}. \textit{Third row:} The geometric phase of the closed path, bounded in the range $[-\pi,\pi]$. \textit{Fourth row:} The dynamic phase. \textit{Fifth row:} The global phase on the surface of the Bloch sphere. The units of the colour bars are multiples of $\pi$.}
	\end{center}
	\label{fig:phaseplots} 
\end{figure}
In this section we demonstrate the quantization of the global phase of the closed path, $\omega=\pm2n\pi$ for $n\in\mathbb{Z}$.
Consider the quaternion  
\begin{equation*}
	\hat{\Psi}\;=\;\hat{U}(t)	\hat{\Psi}_0
\end{equation*}
which is the product of a time independent quaternion $\hat{\Psi}_0$ (initial state) and a unitary $\hat{U}(t)$ that is periodic in $t$ such that
$$\hat{U}(0)\;=\;\hat{U}(2n\pi)
\qquad\qquad\qquad\qquad
\dot{\hat{U}}(0)\;=\;\dot{\hat{U}}(2n\pi)$$
The quaternion is developed as per in section~\ref{sec:quaternion}, to give the SO(3) equation of motion 
\begin{equation*}
\dot{\hat{\Psi}}\;=\;\hat{\mathcal{H}}\;\hat{\Psi}
\end{equation*}
For simplicity and without loss of generality we confine our analysis in this section to the Hopf mappings in the right Cayley basis $\hat{\mathcal{I}}_r, \hat{\mathcal{J}}_r, \hat{\mathcal{K}}_r$. For this numerical analysis of the $\mathbb{S}^1$ fibre bundle we make use of a more complex unitary of the form
\begin{subequations}
\label{eq:unitary}
\begin{align}
\textrm{SU(2) picture:}\;\;\;
&\hat{U}_{\hat{\sigma}}(t)\;=\;
\exp\sqb{-\quat{i}t}
\exp\sqb{\quat{k}\frac{t}{2}}
\exp\sqb{\quat{i}\frac{t}{2}}
\exp\sqb{-\quat{j}t}
\exp\sqb{-\quat{i}t}\\
\textrm{SO(3) picture:}\;\;\;\;
&\hat{U}(t)\;=\;
\exp\sqb{-\hat{\pi}_{i}2t}
\exp\sqb{\hat{\pi}_{k}t}
\exp\sqb{\hat{\pi}_{i}t}
\exp\sqb{-\hat{\pi}_{j}2t}
\exp\sqb{-\hat{\pi}_{i}2t}
\end{align}
\end{subequations}
As the stereographic projection of the Hopf maps has no observable affect on the phases of the $\mathbb{S}^1$ fibre bundle, it suffices to consider the $\hat{\mathcal{I}}_r, \hat{\mathcal{J}}_r, \hat{\mathcal{K}}_r$ projections as equivalent and differing only in terms of their initial states. Consequently we examine the unitary \eqref{eq:unitary} for all initial states of the Bloch sphere. 

To calculate the $\mathbb{S}^1$ phases for all initial states we create a meshgrid of the polar angle $\mu\in[0,\pi]$, and azimuthal angle $\nu\in[0,2\pi]$. In figure~\ref{fig:phaseplots}, the Bloch sphere is parametrized according to three different coordinate systems (top row), where the principle axes are respectively $\vec{\sigma}_i, \vec{\sigma}_j, \vec{\sigma}_k$. The phases of the $\mathbb{S}^1$ fibre bundle are coordinate independent.  While the intermediate meshgrid plots may appear different, they correspond to the same meshgrid on $\mathbb{S}^2$, bottom row.

In \ref{app:geometric_phase} we generate these same grids, and allow an unbounded precession of the geometric phase in the Darboux frame. This still results in identical $\mathbb{S}^2$ meshgrids. However, when we utilize the basis of the normalized partial derivatives, using 3 different conventions for the spherical polar coordinate system, the global phase patterns on $\mathbb{S}^2$ no longer match. These plots are provided for reference. Once again, when the geometric phase is confined to $\gamma\in[-\pi,\pi]$, the global phase in all coordinate systems match that shown in figure~\ref{fig:phaseplots}. 
This numerical check is performed for two reasons,
\begin{itemize}
	\item[$(i)$] To verify the definitions of the global, geometric and dynamic phases are equal in all coordinate systems.
	\item[$(ii)$] To verify the clipping of the geometric phase in the range $[-\pi,\pi]$. 
\end{itemize}
Confining the geometric phase to the range $[-\pi,\pi]$ rectifies all concerns, despite the aesthetic appeal of the plots obtained from an unbounded geometric phase precession.

The bottom row of figure~\ref{fig:phaseplots} shows the global phase is quantized by $\pm2n\pi$ for $n\in\mathbb{Z}$. Indeed this is a property of the $\mathbb{S}^1$ fibre bundle for all smooth, continuous, and closed paths of $\mathbb{S}^2$ generated by the unit quaternion.

\section{The $\mathbb{C}^2$ Spinor} 
\label{sec:spinor}

Presently we move our focus to the mathematical treatment of the  unit quaternion typical of the literature in Quantum Mechanics. The  $\mathbb{C}^2$ spinor is a unit quaternion \cite{Wharton:15} and is known as \textit{`the Qubit'} in the Quantum theory. This discussion is confined to the SU(2) picture, such that $\hat{\Psi}_{\hat{\sigma}},\hat{U}_{\hat{\sigma}},\hat{\mathcal{H}}_{\hat{\sigma}}\in\;$SU(2).
Beginning from the quaternion equation of motion
\begin{equation*}
	\dot{\hat{\Psi}}_{\hat{\sigma}}\;=\;\hat{\mathcal{H}}_{\hat{\sigma}}\;\hat{\Psi}_{\hat{\sigma}}
\end{equation*}
we work backward to the representation found in modern Quantum Mechanics. 
First expand the above in matrix form.
\begin{equation*}
\begin{pmatrix*}
\dot{\alpha} & -\dot{\beta}^* \\ 
\dot{\beta} & \dot{\alpha}^*
\end{pmatrix*}
\;=\;\frac{1}{2}
\begin{pmatrix*}
i\mathcal{H}^i & \mathcal{H}^j + i\mathcal{H}^k \\ 
-\mathcal{H}^j + i\mathcal{H}^k & -i\mathcal{H}^i
\end{pmatrix*}
\begin{pmatrix*}
\alpha & -\beta^* \\ 
\beta & \alpha^*
\end{pmatrix*}
\end{equation*}
with $(\alpha,\beta)=(q_1+iq_i,-q_j+iq_k)$.
Factor out $-i$
\begin{equation*}
\begin{pmatrix*}
	\dot{\alpha} & -\dot{\beta}^* \\ 
	\dot{\beta} & \dot{\alpha}^*
\end{pmatrix*}
\;=\;\frac{-i}{2}
\begin{pmatrix*}
	-\mathcal{H}^i & i\mathcal{H}^j -\mathcal{H}^k \\ 
	-i\mathcal{H}^j - \mathcal{H}^k & \mathcal{H}^i
\end{pmatrix*}
\begin{pmatrix*}
	\alpha & -\beta^* \\ 
	\beta & \alpha^*
\end{pmatrix*}
\end{equation*}
We denote
\begin{align*}
H^z\;=\;-\mathcal{H}^i
\qquad\qquad\qquad
H^y\;=\;-\mathcal{H}^j
\qquad\qquad\qquad
H^x\;=\;-\mathcal{H}^k
\end{align*}
to find
\begin{equation*}
\begin{pmatrix*}
\dot{\alpha} & -\dot{\beta}^* \\ 
\dot{\beta} & \dot{\alpha}^*
\end{pmatrix*}
\;=\;\frac{-i}{2}
\begin{pmatrix*}
H^z & -iH^y + H^x \\ 
iH^y + H^x & -H^z
\end{pmatrix*}
\begin{pmatrix*}
\alpha & -\beta^* \\ 
\beta & \alpha^*
\end{pmatrix*}
\end{equation*}
We denote
\begin{align*}
\ket{\psi^+}\;&=\;
\begin{pmatrix}
	\alpha \\ \beta 
\end{pmatrix}\qquad\qquad\qquad\qquad
\ket{\psi^-}\;=\;
\begin{pmatrix}
	-\beta^* \\ \alpha^*
\end{pmatrix}
\end{align*}
where $\ket{\psi^\pm}\in\mathbb{C}^2$ are spinors. Plugging in we find
\begin{equation*}
\begin{pmatrix*}
\ket{\dot{\psi}^+} & \ket{\dot{\psi}^-}
\end{pmatrix*}
\;=\;\frac{-i}{2}
\begin{pmatrix*}
H^z & -iH^y + H^x \\ 
iH^y + H^x & -H^z
\end{pmatrix*}
\begin{pmatrix*}
\ket{\psi^+} & \ket{\psi^-}
\end{pmatrix*}
\end{equation*}
We ascribe the Pauli matrices
\begin{align*}
\quat{z}\;&=\;
\begin{pmatrix}
	1 & 0 \\ 
	0 & -1
\end{pmatrix}\;=\;-i\quat{i}	
\\
\quat{y}\;&=\;
\begin{pmatrix}
	0 & -i \\ 
	i & 0
\end{pmatrix}\;=\;-i\quat{j}
\\
\quat{x}\;&=\;
\begin{pmatrix}
	0\; & \;1 \\ 
	1\; & \;0
\end{pmatrix}\;=\;-i\quat{k}
\end{align*}
to define
\begin{equation*}
\hat{H}\;=\;\frac{H^x}{2}\quat{x}+\frac{H^y}{2}\quat{y}+\frac{H^z}{2}\quat{z}
\end{equation*}
and find
\begin{equation*}
\ket{\dot{\psi}^\pm} 
\;=\;-i\hat{H}\ket{\psi^\pm} 
\end{equation*}
Multiply both sides by $i$ to arrive at the Schr\"odinger equation
\begin{equation*}
i\ket{\dot{\psi}^\pm} 
\;=\;\hat{H}\ket{\psi^\pm} 
\end{equation*}
Define the spin half basis
\begin{equation*}
\ket{\uparrow\;}\;=\;\binom{1}{0}\qquad\qquad\qquad\ket{\downarrow\;}\;=\;\binom{0}{1}
\end{equation*}
Expand the $\ket{\psi^+}$ spinor as
\begin{equation*}
\ket{\psi^+}\;=\;\alpha\;\ket{\uparrow\;}+\beta\;\ket{\downarrow\;}
\end{equation*}
The magnitude of the complex numbers are interpreted as probability amplitudes that give the probability of a measured result being either spin-up or spin-down. 
$$P_\uparrow\;=\;|\braket{\;\uparrow}{\psi^+}|^2\;=\;
|\alpha|^2\qquad\qquad\qquad P_\downarrow\;=\;|\braket{\;\downarrow}{\psi^+}|^2\;=\;|\beta|^2$$
This is the Born rule applied to the $\mathbb{C}^2$ spinor (unit quaternion) to describe particle spin. 
This perspective is justified by the Copenhagen Interpretation of modern Quantum Mechanics, as it states that the particle exists in both spin states at the same time until the point of measurement, when the superposition collapses to return the measured value 
with a probability $P_\uparrow$ for spin-up and $P_\downarrow$ for spin-down.
This is an axiom of Quantum Information theory, and Quantum Computing, and also finds application in the measurement of position and energy in modern Quantum Mechanics, as surmised in \ref{app:born}.

The analysis of sections~\ref{sec:quaternion} and \ref{sec:hopf} is in conflict with the analysis above, and demonstrates the interpretation of the square magnitude of the complex numbers as a probability measure is a mathematical error, and persists due to a gross misunderstanding of quaternion algebra, and complex analysis. \textit{``The appearance of probability is merely an expression of our ignorance of the true variables in terms of which one can find casual laws''}\cite[{\it page} 114]{Bohm:89}[David Bohm]. Considering the modern theory of quantum spin in this light, it is shown to be wholly incorrect as the true casual laws are those derived in this article from quaternion algebra, and the true variables are those described by the unit quaternion, the $\mathbb{S}^2$ dynamics and the $\mathbb{S}^1$ fibre bundle.

\section{Conclusions and Outlook}

The contrasting accounts of quaternion algebra found in sections~\ref{sec:quaternion} and \ref{sec:hopf}, relative to the traditional treatment of the quaternion via the Born rule and $\mathbb{C}^2$ spinor, section~\ref{sec:spinor}, demonstrates the  widespread misunderstanding of quaternion algebra inherent in modern Quantum Mechanics. It is shown the fundamental groups of Classical Mechanics and Quantum Mechanics, SO(3) and SU(2), are isomorphic. These are equivalent representations of the kinematics of 3D space, that employ different algebraic structures of the unit quaternion, section~\ref{sec:quaternion}. A loss of information in the 6 Hopf mappings from $\mathbb{S}^3$ to $\mathbb{S}^2$, is visible via the stereographic projections \eqref{eq:hopf_project} illustrated in figure~\ref{fig:fiberbundle}. Information is retained in the 3D kinematics via the $\mathbb{S}^1$ fibre bundle, and the global phase is quantized in integer multiples of $2\pi$, section~\ref{sec:hopf}. Whilst the Hopf Fibration is a well known fundamental example of a fibre bundle, the 6 Hopf maps and the closed form definitions of the global, geometric and dynamic phases of the $\mathbb{S}^1$ fibre bundle, are new additions to the literature. 

The Born rule is an observer dependent collapse of the wave-function - which presupposes that the fundamental particles exist in a superposition of spin states prior to measurement. It negates the role of the complex numbers and the quaternion to describe rotations in 2 and 3-dimensional spaces. Notwithstanding, the magnitude of a complex number is a length measure, and has no correspondence with probability whatsoever. These are the axioms upon which the field of Quantum Information Theory and Quantum Computing is founded. A more plausible account for the intrinsic spin of the fundamental particles can be found in the $\mathbb{S}^1$ fibre bundle of the Hopf Fibration. The global phase is a measure of the total anholonomy and is quantized for closed $\mathbb{S}^2$ paths. The global phase determines whether a full rotation in $\mathbb{S}^2$ corresponds to a full or half rotation in $\mathbb{S}^3$, lending to the suggestion that the magnetic moment is 4 dimensional, and the intrinsic spin observed in 3D is a Hopf projection (a shadow) of the multi-dimensional particle. 

The existence of higher dimensional spaces is not a new concept in Physics, as one only needs to look to the theoretical studies of observed phenomena in the Large Hadron Collider at CERN, which are largely attributed to the SU(3) group and the $\mathbb{C}^3$ spinor - an even higher dimensional group than the quaternion. Furthermore one can reason that should the entangled state be faithfully described by the $\mathbb{C}^4$ spinor (an even higher dimensional space again), then the observation of two spatially separated particles is an illusion, derived from looking at a multi-dimensional object in 3D. These sentiments are not new and have been stated before, {\it ``the guiding wave, in the general case, propagates not in ordinary three-space but in a multi-dimensional configuration space is the origin of the notorious `non-locality' of Quantum Mechanics''}\cite[{\it ch} 14]{Bell:87} [John Bell]. 

This analysis rouses a natural first question on the mind of any scientist with a curiosity on the quantum theory: \textit{`Do these results provide a mechanism to account for the Bell Inequality results? As these inequalities are supposed to rule out hidden variables accounts for Quantum Mechanics under reasonable assumptions.'} The formalism presented herein corresponds to the single qubit case - whereas the Bell Inequalities would consider correlations between two qubits, the bipartite state. To properly address this question, the current analysis would need to be extended to the $\mathbb{C}^4$ spinor, and include an account of the SU(4) group, and incorporate the higher dimensional Hopf mappings of $\mathbb{S}^7\mapsto\mathbb{S}^4$ and $\mathbb{S}^{15}\mapsto\mathbb{S}^8$ \cite{Adams:60}. In a general sense the concepts discussed in this article can be conceptually expressed in terms of Abbott's 2D flatland world - where a banana suspended in 3D space would intersect the 2D plane of the flatland world creating two spatially separated discs. To the flatlanders, these discs would represent two particles in an entangled state, and their experiments would reveal both non-local, and faster than light correlations between the spatially separated particles. 

This article refutes the Born rule, and the Copenhagen Interpretation of Quantum Mechanics, which is the probability interpretation of the complex numbers. \textit{``God knows I am no friend of probability theory, I have hated it since the first moment our dear friend Max Born gave it birth.''}\cite{Moore:89} \textit{``I am not opposing a few special statements of Quantum Mechanics held today (1950s). I am opposing as it were the whole of it, I am opposing it's basic views that have been shaped 25 years ago, when Max Born put forward his probability interpretation which was accepted by almost everybody.''}\cite{Schrodinger:95}  [Erwin Schr\"odinger]. This view is shared among the founding fathers of Quantum Mechanics. Einstein has himself stated: \textit{``Quantum Mechanics is certainly imposing. But an inner voice tells me that this is not yet the real thing. The theory says a lot but does not bring us any closer to the secrets of the `Old One'. I, at any rate, am convinced that He is not playing at dice.''}\cite{Einstein:26}. In parallel with the probability interpretation of the complex numbers there exists the measurement problem of Quantum Mechanics, the notion of the quantum particle existing in mutually opposed states until the point of measurement, and the barrier between the quantum and classical realms: As Schr\"odinger emphatically states: \textit{``The world is given to me only once, not one existing and one perceived. Subject and object are only one. The barrier between them cannot be said to be broken down as a result of recent experience in the physical sciences, for this barrier does not exist.''}\cite{Schrodinger:56} John Bell has similarly expressed the same: \textit{``Nobody knows where the boundary between the Quantum and Classical domain is situated. More plausible to me is that we will find there is no boundary.''}\cite[ch 4]{Bell:87} 

These and related statements made by the founding fathers of Quantum Mechanics are echoed in this article, which is presented as an affirmation of sentiments of concern regarding the probability interpretation of the square magnitude of the complex numbers. The fundamental algebra of Quantum Mechanics, SU(2), is here demonstrated isomorphic to the fundamental algebra of Classical Mechanics, SO(3). Quantum Mechanics has failed to acknowledge this. The magnitude of the complex numbers is a length measure, and given that this is a mathematical truth, then the concept of a Quantum-Classical boundary is misguided. Deterministic equations have only become non-deterministic due to a mathematical \textit{faux pas}. The measurement problem of modern Quantum Mechanics is brought about by first of all negating information available in higher dimensional spaces via the Born rule, and then entertaining wonder and debate as to where the hidden variables might have disappeared to - or be found. Indeed these results suggest that the scientific community has found itself in a position reminiscent of fabled stories on the Human condition such as \textit{The Emperor Has No Clothes}, as it would appear that people are unable to see the woods for the trees. The premise of this article is to propose the dissolution of the Born rule and the Copenhagen Interpretation of Quantum Mechanics. The integrity of Quantum Information theory is challenged, as is any realistic possibility of Quantum Computing ever being realised, since the axioms on which these theories are founded are in error.

To conclude it is useful to remind the scientific community that the viewpoints expressed here are not new. They have been stated in different ways by reputable scientists since the birth of modern Quantum Mechanics. While the Quantum theory earned it's name from the quantization of energy, space, and time at the smallest scales, modern Quantum Mechanics emerged as a non-deterministic science founded on the Born rule and Copenhagen Interpretation, which are axioms contrary to the basis of any reasonable scientific investigation. As stated by Max Planck: \textit{``The assumption of an absolute determinism is the essential foundation of every scientific enquiry''}\cite{Heilbron:86}. Put simply, the magnitude of a complex number is a length measure, never a probability measure. \textit{``The quantum hypothesis will eventually find it's exact expression in certain equations which will be a more exact formulation of the law of causality''}\cite{Planck:32}. This article has accounted for the hidden variables of Classical Mechanics,  and has sketched the basic requirements of a multi-dimensional field theory to supersede modern Quantum Mechanics. 

%\section*{Acknowledgements}
%
%Your lies gave me psychosis. Your voodoo brought me to tears. I chose the rope, and wrote a note, while little birdies sang in my ear.

\appendix
%
%%%%%%%%%%%%%%%%%%%%%%%%%%%%%%%%%%%%%%%%%%%%%%%%%%%%%%%%%%%%%%%%%%%%%%%%%%%%%%%%%%%%%%%%%%%%%%%%%%%%%%%%%%%%%%%%%%%%%%%%%%%%%%%%%%%%%%%%%%%%%

\section{The U(1) Gauge}
\label{app:U1gauge}

The simplest example of a principle bundle is the U(1) gauge, the circle group. 
Here we consider the affect of the group transformation,  
\begin{center}
	SU(2) $\;\mapsto\;$	U(1)\;$\times$\;SU(2)
\end{center}
on the geometric phase. 
This gauge transformation of the SU(2) quaternion $\hat{\Psi}_{\hat{\sigma}}$ via $\exp\sqb{\frac{i}{2}\int_0^tdt'\mathcal{H}^1(t')},$ is expressed
\begin{equation*}
	\hat{\Phi}_{\hat{\sigma}}\;=\;\exp\sqb{ \frac{i}{2}\int_0^tdt'\;\mathcal{H}^1(t')} \hat{\Psi}_{\hat{\sigma}}
\end{equation*}
It is immediately observed that the $\mathbb{S}^2$ dynamics of the U(1)\;$\times$\;SU(2) spinor is equivalent to the SU(2) spinor under the 6 Hopf maps of \eqref{eq:hopf_project}. 
We quote the same maps here in the SU(2) cayley basis for completeness.
\begin{subequations}
	\begin{align*}
		\hat{\mathcal{I}}_l\;=\;\hat{\Phi}^\dagger_{\hat{\sigma}}\hat{\sigma}_i\hat{\Phi}_{\hat{\sigma}}
		\;=\;\hat{\Psi}^\dagger_{\hat{\sigma}}\hat{\sigma}_i\hat{\Psi}_{\hat{\sigma}}
		\qquad\qquad\;&
		\hat{\mathcal{I}}_r\;=\;\hat{\Phi}_{\hat{\sigma}}\hat{\sigma}_i\hat{\Phi}^\dagger_{\hat{\sigma}}
		\;=\;\hat{\Psi}_{\hat{\sigma}}\hat{\sigma}_i\hat{\Psi}^\dagger_{\hat{\sigma}}\\
		\hat{\mathcal{J}}_l\;=\;\hat{\Phi}^\dagger_{\hat{\sigma}}\hat{\sigma}_j\hat{\Phi}_{\hat{\sigma}}
		\;=\;\hat{\Psi}^\dagger_{\hat{\sigma}}\hat{\sigma}_j\hat{\Psi}_{\hat{\sigma}}
		\qquad\qquad\;&
		\hat{\mathcal{J}}_r\;=\;\hat{\Phi}_{\hat{\sigma}}\hat{\sigma}_j\hat{\Phi}^\dagger_{\hat{\sigma}}
		\;=\;\hat{\Psi}_{\hat{\sigma}}\hat{\sigma}_j\hat{\Psi}^\dagger_{\hat{\sigma}}\\
		\hat{\mathcal{K}}_l\;=\;\hat{\Phi}^\dagger_{\hat{\sigma}}\hat{\sigma}_k\hat{\Phi}_{\hat{\sigma}}
		\;=\;\hat{\Psi}^\dagger_{\hat{\sigma}}\hat{\sigma}_k\hat{\Psi}_{\hat{\sigma}} 
		\qquad\qquad\;&
		\hat{\mathcal{K}}_r\;=\;\hat{\Phi}_{\hat{\sigma}}\hat{\sigma}_k\hat{\Phi}^\dagger_{\hat{\sigma}} 
		\;=\;\hat{\Psi}_{\hat{\sigma}}\hat{\sigma}_k\hat{\Psi}^\dagger_{\hat{\sigma}} 
	\end{align*}
\end{subequations}
The U(1) gauge cancels out under the Hopf mappings, consequently we can state that the geometric phase is gauge invariant, as the $\mathbb{S}^2$ dynamics of the Bloch vector are equivalent for each map. 
%For completeness, we develop the equation of motion
%\begin{equation*}
%	\dot{\hat{\Phi}}_{\hat{\sigma}}\;=\;
%	\hat{\mathcal{H}}_{\hat{\sigma}}\hat{\Phi}_{\hat{\sigma}}
%\end{equation*}
%with 
%\begin{equation*}
%	\hat{\mathcal{H}}_{\hat{\sigma}}\;=\; 
%	\frac{i\mathcal{H}^1}{2}\hat{\sigma}_1
%	+\frac{\mathcal{H}^i}{2}\hat{\sigma}_i
%	+\frac{\mathcal{H}^j}{2}\hat{\sigma}_j
%	+\frac{\mathcal{H}^k}{2}\hat{\sigma}_k
%\end{equation*}
%and it is seen that the U(1) gauge appears in the $\hat{\sigma}_1$ axis of the Hamiltonian.

\section{Precession of the geometric phase}
\label{app:geometric_phase}

\begin{figure}[t] 
\includegraphics[width=\textwidth]{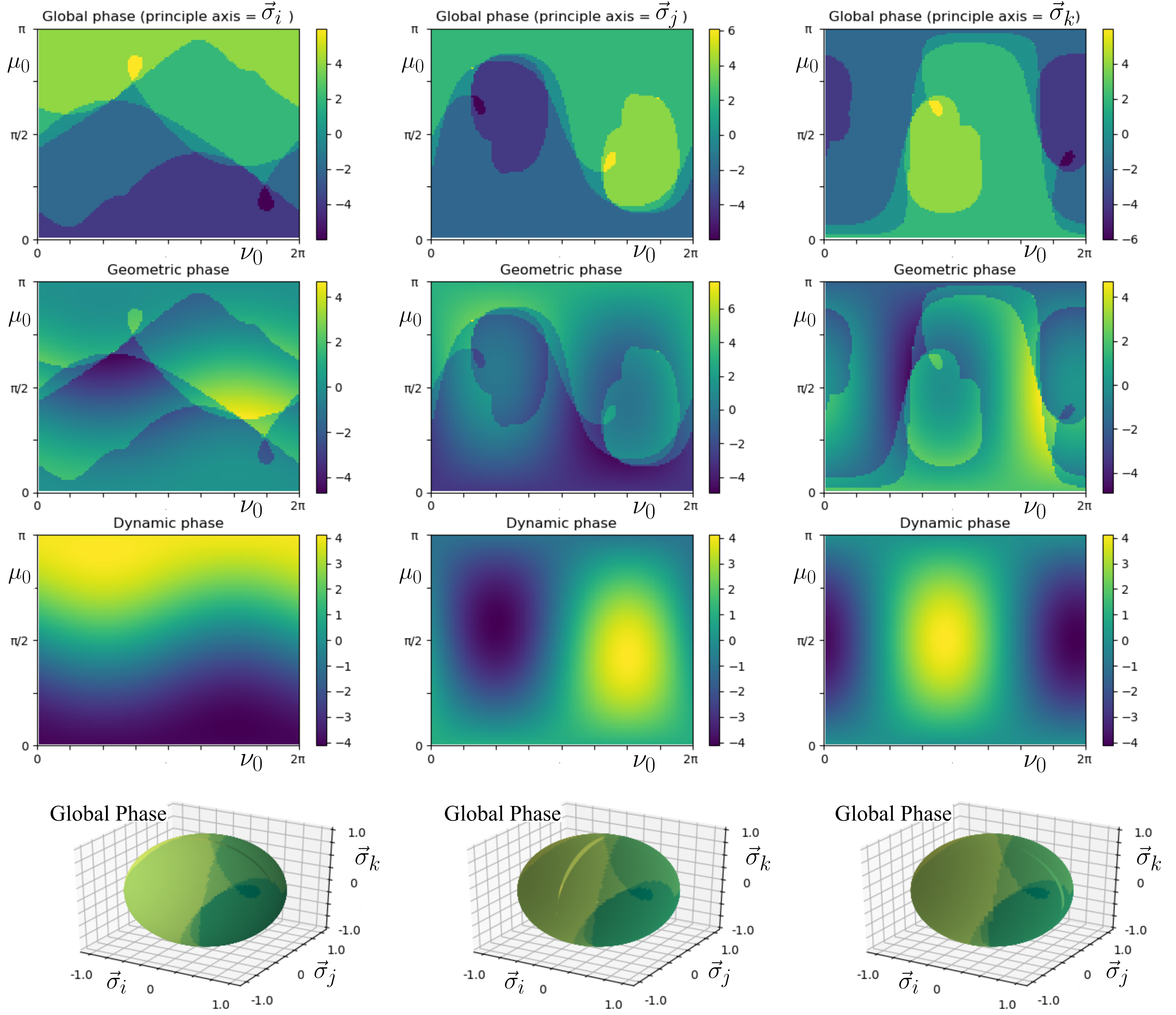} 
\caption{\textit{Top row:} The global phase of the closed path for the unitary of equation~\eqref{eq:unitary}. \textit{Second row:} The unbounded geometric phase of the closed path. \textit{Third row:} The dynamic phase. \textit{Fourth row:} The global phase on the surface of the Bloch sphere. The units of the colour bars are multiples of $\pi$.}
\label{fig:darboux_noclip} 
\end{figure}
\begin{figure}[t]
	\includegraphics[width=\textwidth]{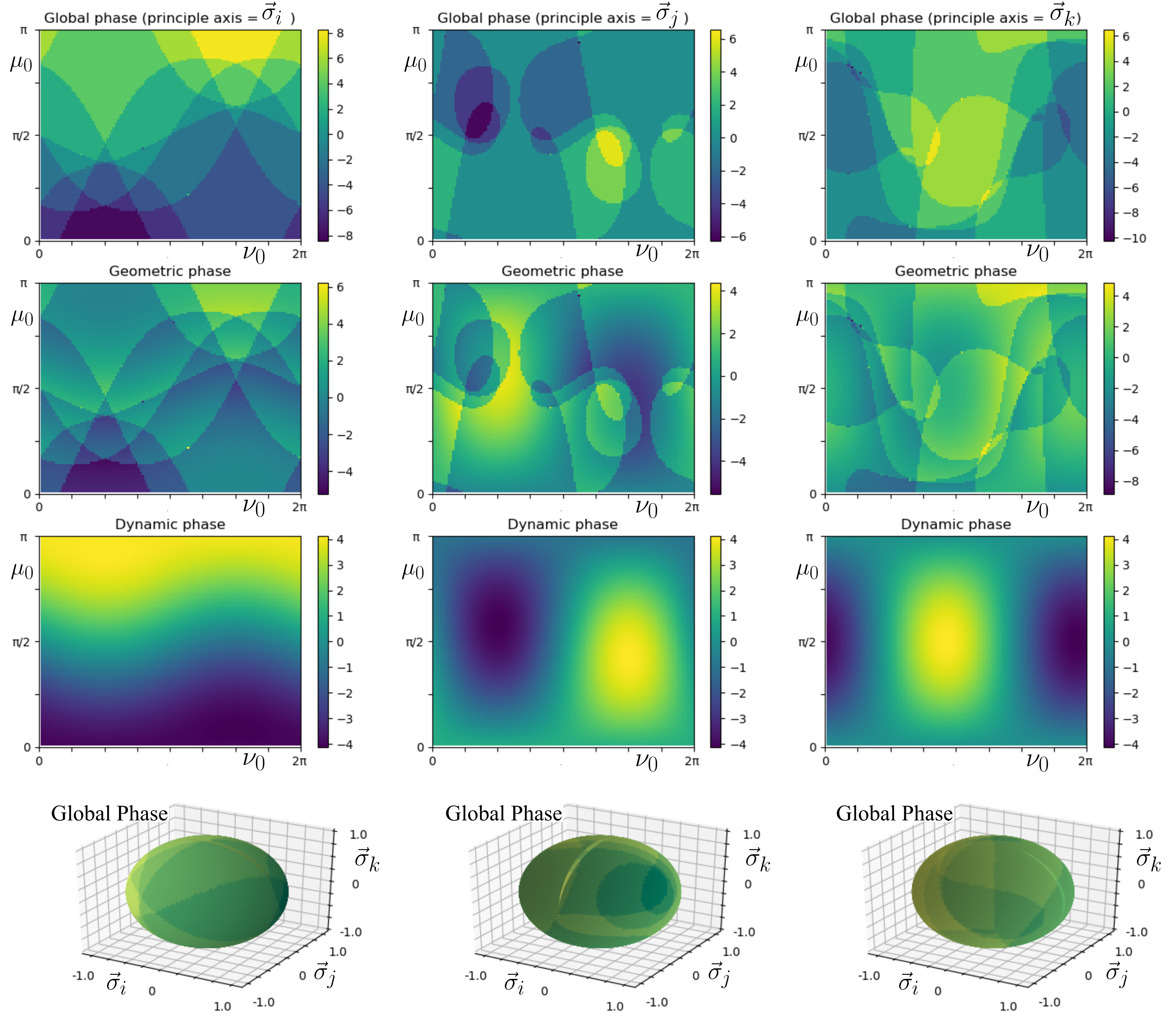}
	\caption{\textit{Top row:} The global phase of the closed path for the unitary of equation~\eqref{eq:unitary}. \textit{Second row:} The unbounded geometric phase of the closed path. \textit{Third row:} The dynamic phase. \textit{Fourth row:} The global phase on the surface of the Bloch sphere. The units of the colour bars are multiples of $\pi$.}
	\label{fig:spherical_polars_noclip} 
\end{figure}

Presented in figure~\ref{fig:darboux_noclip}, for the purposes of reference are the global, geometric and dynamic phases, when the geometric phase is unbounded and allowed to precess in the moving frame taking values outside the limits $[-\pi,\pi]$. Similarly in figure~\ref{fig:spherical_polars_noclip} are the global and geometric phases calculated in the basis of the normalized partial derivatives. Shown are 3 different parametrization conventions in spherical polar coordinates (columns), with principle axes $\cub{\vec{\sigma}_i, \vec{\sigma}_j, \vec{\sigma}_k}$ respectively. When the geometric phase is confined to $[-\pi,\pi]$ the plots of figures~\ref{fig:darboux_noclip} and \ref{fig:spherical_polars_noclip} simplify to those seen in figure~\ref{fig:phaseplots}.

\subsection{Spherical polars with $(\vec{\sigma}_i, \vec{\sigma}_j, \vec{\sigma}_k)$ as the principle axes.}

In spherical polar coordinates, the geometric phase is derived in each coordinate system as follows.
The Bloch vector is parametrized in terms of the polar and azimuthal angles $\cub{\mu,\nu}$:
\begin{align*}
\hat{\sigma}_i\;\textrm{principle axis:}
\qquad\qquad&
\hat{\sigma}_j\;\textrm{principle axis:}
\qquad\qquad\qquad
\hat{\sigma}_k\;\textrm{principle axis:}\\
\vec{\mathcal{R}}\;=\;
\begin{pmatrix}
\cos\cub{\mu}\\ 
\sin\cub{\mu}\cos\cub{\nu}\\ 
\sin\cub{\mu}\sin\cub{\nu}
\end{pmatrix}
\qquad
\vec{\mathcal{R}}\;&=\;
\begin{pmatrix}
\sin\cub{\mu}\sin\cub{\nu}\\ 
\cos\cub{\mu}\\ 
\sin\cub{\mu}\cos\cub{\nu}
\end{pmatrix}
\qquad
\vec{\mathcal{R}}\;=\;
\begin{pmatrix}
\sin\cub{\mu}\cos\cub{\nu}\\ 
\sin\cub{\mu}\sin\cub{\nu}\\
\cos\cub{\mu}
\end{pmatrix}
\end{align*}
The normalized basis of the partial derivatives 
\begin{align*}
\hat{\sigma}_i\;\textrm{principle axis:}
\qquad\qquad
\uve{\mu}\;&=\;
\begin{pmatrix}
	-\sin\cub{\mu}\\ 
	\cos\cub{\mu}\cos\cub{\nu}\\ 
	\cos\cub{\mu}\sin\cub{\nu}
\end{pmatrix}
\qquad\qquad
\uve{\nu}\;=\;
\begin{pmatrix}
	0\\
	-\sin\cub{\nu}\\ 
	\cos\cub{\nu}
\end{pmatrix}\\
\hat{\sigma}_j\;\textrm{principle axis:}
\qquad\qquad
\uve{\mu}\;&=\;
\begin{pmatrix}
	\cos\cub{\mu}\sin\cub{\nu}\\ 
	-\sin\cub{\mu}\\ 
	\cos\cub{\mu}\cos\cub{\nu}
\end{pmatrix}
\qquad\qquad 
\uve{\nu}\;=\;
\begin{pmatrix}
	\cos\cub{\nu}\\ 
	0\\ 
	-\sin\cub{\nu}
\end{pmatrix}\\
\hat{\sigma}_k\;\textrm{principle axis:}
\qquad\qquad
\uve{\mu}\;&=\;
\begin{pmatrix}
	\cos\cub{\mu}\cos\cub{\nu}\\ 
	\cos\cub{\mu}\sin\cub{\nu}\\
	-\sin\cub{\mu}
\end{pmatrix}
\qquad\qquad 
\uve{\nu}\;=\;
\begin{pmatrix}
	-\sin\cub{\nu}\\ 
	\cos\cub{\nu}\\
	0
\end{pmatrix}
\end{align*}
From the equation of parallel transport:
\begin{equation*}
	\frac{D\vec{\mathcal{V}}}{Dt}\;=\;\dot{\vec{\mathcal{V}}}\cdot\uve{\lambda}\;=\;0
	\qquad\qquad \forall\;\; \uve{\lambda}\in[\uve{\mu},\uve{\nu}]
\end{equation*}
we obtain the simultaneous equations
\begin{equation*}
\frac{D\vec{\mathcal{V}}}{Dt}\;=\;
\begin{cases}
\dot{\mathcal{V}}^\mu + \dot{e}_{\nu}\cdot\uve{\mu} \;\mathcal{V}^\nu \;=\;0 \\ 
\dot{\mathcal{V}}^\nu + \dot{e}_{\mu}\cdot\uve{\nu} \;\mathcal{V}^\mu \;=\;0
\end{cases}
\end{equation*}
The derivative of the geometric phase is defined $\dot{\gamma}\equiv\dot{e}_{\mu}\cdot\uve{\nu}=-\dot{e}_{\nu}\cdot\uve{\mu}$.
Express the equation of parallel transport in matrix form
\begin{equation*}
\begin{pmatrix}
\dot{\mathcal{V}}^\mu \\ \dot{\mathcal{V}}^\nu
\end{pmatrix}\;=\; 
\begin{pmatrix}
0 & \dot{\gamma} \\
-\dot{\gamma} & 0
\end{pmatrix}
\begin{pmatrix}
\mathcal{V}^\mu \\ \mathcal{V}^\nu 
\end{pmatrix}
\end{equation*}
and the tangent vector evolves from it's initial state according to
\begin{equation*} 
\begin{pmatrix}
\mathcal{V}^\mu(t) \\ 
\mathcal{V}^\nu(t)
\end{pmatrix}
\;=\;
\begin{pmatrix}
\cos\cub{\gamma} & \sin\cub{\gamma} \\
-\sin\cub{\gamma} & \cos\cub{\gamma} 
\end{pmatrix}
\begin{pmatrix}
\mathcal{V}^\mu(0) \\ 
\mathcal{V}^\nu(0)
\end{pmatrix}
\end{equation*}
In these defined coordinate systems the geometric phase is respectively:
\begin{equation*}
	\gamma\;=\;\int_0^tdt'\;\dot{\gamma}(t')
\end{equation*} 
with
\begin{align*}
\hat{\sigma}_i\;\textrm{principle axis:}
\qquad\qquad
\dot{\gamma}\;&\equiv\; 
-\mathcal{H}^i\mathcal{R}^i+\cub{\frac{\mathcal{H}^j\mathcal{R}^j+\mathcal{H}^k\mathcal{R}^k}
{\cub{\mathcal{R}^j}^2+\cub{\mathcal{R}^k}^2}}\cub{\mathcal{R}^i}^2
\\
\hat{\sigma}_j\;\textrm{principle axis:}
\qquad\qquad
\dot{\gamma}\;&\equiv\; 
-\mathcal{H}^j\mathcal{R}^j+\cub{\frac{\mathcal{H}^i\mathcal{R}^i+\mathcal{H}^k\mathcal{R}^k}
{\cub{\mathcal{R}^i}^2+\cub{\mathcal{R}^k}^2}}\cub{\mathcal{R}^j}^2
\\
\hat{\sigma}_k\;\textrm{principle axis:}
\qquad\qquad
\dot{\gamma}\;&\equiv\; 
-\mathcal{H}^k\mathcal{R}^k+\cub{\frac{\mathcal{H}^i\mathcal{R}^i+\mathcal{H}^j\mathcal{R}^j}
{\cub{\mathcal{R}^i}^2+\cub{\mathcal{R}^j}^2}}\cub{\mathcal{R}^k}^2
\end{align*}

\section{Measurement of spatial and energetic modes via the Born rule}
\label{app:born}

The energy eigenstates  of the one dimensional harmonic oscillator are the normalized Hermite functions\footnote[7]{Their analytic form is unquoted here as this discussion is for illustrative purposes only.} denoted $\chi_n(x)$ for $n=0,1,2,\dots,$. The eigenstates form an orthonormal basis such that 
\begin{equation*}
	\int_{-\infty}^\infty dx \; \chi_n(x)\chi_m(x)\;=\;	
	\int_{-\infty}^\infty dx \; \braket{\chi_n}{x}\braket{x}{\chi_m}\;=\;
	\braket{\chi_n}{\chi_m}\;=\;\delta_{nm}
\end{equation*}
These relations show the equivalence of Dirac's bra-ket notation, and standard notation for functions, since $\braket{x}{\chi_m}=\chi_m(x)$.
The corresponding energy levels are eigenvalues of $\hat{H}\ket{\chi_n}=E_n\ket{\chi_n}$ with $\hat{H}$ being the Hamiltonian operator of Schr\"odinger's equation, and
\begin{equation*}
	E_n\;=\;\hbar\omega\cub{n+\tfrac{1}{2}}
\end{equation*}
The wave-function is expanded as the linear superposition
\begin{equation*}
	\ket{\Phi}\;=\;\sum_{n=0}^{N-1} c_n\ket{\chi_n}
\end{equation*}
where $c_n\in\mathbb{C}$ and the wave-function is normalized such that 
\begin{equation*}
	\sum_{n=0}^{N-1} |c_n|^2\;=\;1
\end{equation*}
Consequently the (uncoupled) wave-function evolves in time as 
\begin{equation}
	\label{eq:wavefn}
	\ket{\Phi(t)}\;=\;\sum_{n=0}^{N-1} c_n e^{-iE_nt}\ket{\chi_n}
\end{equation}
Due to the orthogonality of this system of equations we have
\begin{equation*}
	\int_{-\infty}^{\infty}dx\;\braket{\Phi(t)}{x}\braket{x}{\Phi(t)}\;=\;	\braket{\Phi(t)}{\Phi(t)}\;=\;1
\end{equation*}
for all times $t$.
From here there are typically two means by which the Born rule is applied to the wave-function $\ket{\Phi}$.
\begin{enumerate}
	\item Energy eigenstate: the probability of finding the wave-function in an eigenstate $\ket{\chi_m}$ is 
	$$\hat{P}_{\ket{\chi_m}}\;=\;\braket{\Phi(t)}{\chi_m}\braket{\chi_m}{\Phi(t)}\;=\;|c_m|^2$$ 
	\item Position: the probability of finding a particle in the spatial interval $x_{ab}=(x_a,x_b)$ is the integral
	$$P_{x_{ab}}\;=\;\int_{x_a}^{x_b}dx\;\braket{\Phi(t)}{x}\braket{x}{\Phi(t)}$$ 
\end{enumerate}
The assignment of the square magnitude of the complex number to describe the probable result of measurement is contrary to the traditional role of the complex numbers - which are to describe rotations in the 2-dimensional plane. These axioms are also contrary to the traditional definition of probability - which is a statistical distribution of deterministic states. Should these considerations taken into account the `one dimensional' wave-function of equation~\eqref{eq:wavefn}, would describe a rotation in a complex multi-dimensional space, rather than a probabilistic distribution in a one dimensional space.

%\begin{figure}[h] 
%	\includegraphics[width=\textwidth]{figures/hopf_gene.png} 
%	\caption{\textit{Left:} Illustration of the Hopf Fibration.
%	\textit{Right:} Gene Wilder [1933-2016] appreciates the wonders of the human mind.} 
%\end{figure}

\end{document}